\documentclass{IEEEaccess}
\makeatletter
\@whilenum\allocationnumber<88\do{\newbox\accessreservedbox}
\makeatother
\usepackage{cite}
\usepackage{amsmath,amssymb,amsfonts}
\usepackage{algorithmic}
\usepackage{graphicx}
\usepackage{textcomp}

\usepackage{bm}
\makeatletter
\AtBeginDocument{\DeclareMathVersion{bold}
\SetSymbolFont{operators}{bold}{T1}{times}{b}{n}
\SetSymbolFont{NewLetters}{bold}{T1}{times}{b}{it}
\SetMathAlphabet{\mathrm}{bold}{T1}{times}{b}{n}
\SetMathAlphabet{\mathit}{bold}{T1}{times}{b}{it}
\SetMathAlphabet{\mathbf}{bold}{T1}{times}{b}{n}
\SetMathAlphabet{\mathtt}{bold}{OT1}{pcr}{b}{n}
\SetSymbolFont{symbols}{bold}{OMS}{cmsy}{b}{n}
\renewcommand\boldmath{\@nomath\boldmath\mathversion{bold}}}
\makeatother

\def\BibTeX{{\rm B\kern-.05em{\sc i\kern-.025em b}\kern-.08em
    T\kern-.1667em\lower.7ex\hbox{E}\kern-.125emX}}

\usepackage{pifont}
\newcommand{\cmark}{\ding{51}}

\usepackage[ruled, linesnumbered]{algorithm2e}
\usepackage{booktabs}
\usepackage{url}
\usepackage{subcaption}
\usepackage{multirow}
\usepackage{siunitx}
\usepackage{flafter}
\usepackage{orcidlink}
\usepackage{tikz}
\NewSpotColorSpace{PANTONE}
\AddSpotColor{PANTONE} {PANTONE3015C} {PANTONE\SpotSpace 3015\SpotSpace C} {1 0.3 0 0.2}
\SetPageColorSpace{PANTONE}
\usepackage{xcolor}
\newcommand*\circled[1]{\tikz[baseline=(char.base)]{
  \node[shape=circle,fill,inner sep=2pt] (char)
{\textcolor{white}{\textbf{#1}}};}}

\sloppypar

\newcommand{\secref}[1]{§\ref{#1}}
\newcommand{\figref}[1]{Figure~\ref{#1}}

\newcommand{\bdm}{BioDynaMo}
\newcommand{\ta}{TeraAgent}

\newcommand{\etal}{et al.}
\newcommand{\head}[1]{\noindent\textbf{#1.}}
\newcommand{\result}[1]{#1}

\newcommand{\ESMaxAgentsExact}{501.51 billion}
\newcommand{\ESMaxAgentsInTrillion}{half a trillion}
\newcommand{\ESNumCPUs}{84,096}
\newcommand{\ESNumNodes}{438}
\newcommand{\ESMemoryConsumptionInTB}{92}
\newcommand{\ESRuntimePerSecond}{147}
\newcommand{\ESFactorImprovementSOTA}{292}
\newcommand{\ESFactorImprovementLiterature}{84}
\newcommand{\ESAgentsBDM}{1.72 billion}
\newcommand{\ESAgentsLiterature}{6 billion}

\newcommand{\ESMidAgents}{102.4 billion}
\newcommand{\ESMidNumNodes}{128}
\newcommand{\ESMidNumCPUs}{24,576}
\newcommand{\ESMidMemoryConsumptionInTB}{40}
\newcommand{\ESMidRuntimePerSecond}{7.1}

\newif\ifaccessdraft
\accessdrafttrue

\ifaccessdraft
\usepackage{todonotes}
\usepackage[normalem]{ulem}

\newcommand{\agyadelcite}[2][]{[..]}
\newcommand{\agyadel}[1]{{\color{gray}\let\cite\agyadelcite\sout{#1}}}
\newcounter{agyabox}
\newcommand{\agyacomment}[1]{\refstepcounter{agyabox}\todo[inline, color=white, bordercolor=red, size=\footnotesize]{\textcolor{red}{\textbf{[R\theagyabox] Giray:} #1}}}
\definecolor{agyagreen}{RGB}{0,120,0}
\newcommand{\agyaccomment}[1]{\todo[inline, color=white, bordercolor=agyagreen, size=\footnotesize]{\textcolor{agyagreen}{\textbf{Opus (makes sense to Giray):} #1}}}
\else

\newcommand{\agyadel}[1]{}
\newcommand{\agyacomment}[1]{}
\newcommand{\agyaccomment}[1]{}
\renewcommand{\agycomment}[1]{}
\renewcommand{\mscomment}[1]{}
\fi

\newif\ifthesis
\thesisfalse
\newcommand{\ThesisPaper}[2]{\ifthesis#1\else#2\fi}

\begin{document}
\history{Received XX Month 2026; accepted XX Month 2026; date of publication XX Month 2026; date of current version XX Month 2026.}
\doi{10.1109/ACCESS.2026.XXXXXXX}

\title{\ta{}: A Distributed Agent-Based Simulation Engine\\for Simulating Half a Trillion Agents}

\author{\uppercase{Lukas Breitwieser}\orcidlink{0000-0003-2265-8615}\authorrefmark{1}, 
\uppercase{Ahmad Hesam}\orcidlink{0000-0001-7331-1000}\authorrefmark{2},
\uppercase{A. G{{i}}ray Ya\u{g}l{i}k{\c{c}}{i}}\orcidlink{0000-0002-9333-6077}\authorrefmark{3},
\uppercase{Mohammad Sadrosadati}\orcidlink{0000-0002-4029-0175}\authorrefmark{1}, 
\uppercase{Fons Rademakers}\orcidlink{0000-0002-3571-9635}\authorrefmark{4}, 
AND
\uppercase{Onur Mutlu}\orcidlink{0000-0002-0075-2312}\authorrefmark{1}}

\address[1]{ETH Zurich, Switzerland}
\address[2]{Delft University of Technology, The Netherlands}
\address[3]{CISPA, Germany}
\address[4]{CERN, Switzerland}

\corresp{Corresponding author: Lukas Breitwieser (e-mail: lukas.breitwieser@gmail.com).}

\begin{keywords}
Agent-based modeling,
delta encoding,
distributed computing,
high-performance computing,
high-performance simulation, 
MPI, 
performance optimization, 
scalability, 
serialization.
\end{keywords}

\newcommand{\cognoLungInjury}[0]{cogno_mechanistic_2024,cogno_thesis,cogno_agent-based_2022,cogno_3d_2022}
\newcommand{\abmImmuneSystem}[0]{montealegre_agent-based_2012,shinde_review_2018,bauer2009,truszkowska_predicting_2022,perrin_agent-based_2006,pappalardo_agent_2018,folcik_using_2011,folcik_basic_2007,chiacchio_agent-based_2014,jacob_swarm-based_2011}
\newcommand{\abmChronicDiseases}[0]{li2016,montagna_agent-based_2017,nianogo_agent-based_2015,schryver_emulating_2015,mi_agent-based_2007,khademi_agent-based_2018,li_assessing_2014,archbold_agent-based_2024,giabbanelli_application_2021,morshed_systematic_2019,beheshti_comparing_2017,giabbanelli_using_2017,squires_long-term_2023}
\newcommand{\abmHealthcareSystemOpt}[0]{cassidy_mathematical_2019,cabrera_simulation_2012,ajmi_agent-based_2019}
\newcommand{\abmPharmacoDynamics}[0]{gao_developing_2017}
\newcommand{\abmCardiovasularRiskFactors}[0]{li2018}

\newcommand{\abmMedicine}[0]{\cognoLungInjury,\abmImmuneSystem,\abmChronicDiseases,\abmHealthcareSystemOpt,\abmPharmacoDynamics,\abmCardiovasularRiskFactors}

\newcommand{\abmCancer}[0]{demontigny_2021,duswald_2024,cogno_biomedicin_2024,metzcar_review_2019,wang2015simulating,olsen_multiscale_2013,wang_multi-scale_2013,an_agent-based_2015,wang_integrated_2015,poleszczuk_agent-based_2016,norton_multiscale_2019,heidary_double-edged_2020,rivera_agent-based_2022,van_genderen_agent-based_2024}
\newcommand{\abmNeuronGrowth}[0]{zublerdouglas2009framework,duswald2024calibrationstochasticagentbasedneuron,torben-nielsen_context-aware_2014,de_schutter_efficient_2023}
\newcommand{\abmCorticalLamination}[0]{abubacar_neuronal_growth,zublerdissertation}
\newcommand{\abmNervousSystem}[0]{\abmNeuronGrowth,\abmCorticalLamination,pennisi2015,caffrey_silico_2014,avin_agent-based_2021}
\newcommand{\abmMorphoGenesis}[0]{glen2019,tang_phenotypic_2011,lambert_bayesian_2018,walpole_agent-based_2015,camacho-gomez_3d_2022,dalle_nogare_netlogo_2020,bonabeau_classical_1997,thorne_agent-based_2007}
\newcommand{\abmBiofilms}[0]{kreft2001,wilmoth_microfluidics_2018,li_agent-based_2024,latif_multiscale_2018,sweeney_agent-based_2019,nagarajan_agent-based_2022,koshy-chenthittayil_agent_2021}
\newcommand{\abmBiologyOther}[0]{soheilypour_agent-based_2018,an_agent-based_2009,griffin_agent-based_2006,gorochowski_agent-based_2016}

\newcommand{\abmBiology}[0]{\abmCancer,\abmNervousSystem,\abmMorphoGenesis,\abmBiofilms,\abmBiologyOther}

\newcommand{\abmCOVID}[0]{hesam_2024,ozik_citycovid_2021,yin_data_2021,kou_multi-scale_2021,faucher_agent-based_2022,pescarmona_agent-based_2021,cattaneo_agent-based_2022,kerr_covasim_2021,truszkowska_exploring_2023,hoertel_facing_2020,kumaresan_fitting_2023,ogden_mathematical_2024,topirceanu_impact_2023}
\newcommand{\abmInfluenza}[0]{laskowski_agent-based_2011,depasse_does_2017,karimi_effect_2015,kumar_policies_2013}
\newcommand{\abmMalaria}[0]{gharakhanlou_spatial_2020,pizzitutti_validated_2015,modu_agent-based_2023,bomblies_agent-based_2014,smith_agent-based_2018}
\newcommand{\abmHIV}[0]{teweldemedhin_agent-based_2004,brookmeyer_combination_2014,anderle_modelling_2024}
\newcommand{\abmPublicHealthInterventions}[0]{getz_agent-based_2019,liu_role_2015}
\newcommand{\abmEpidemicForecasting}[0]{tabataba_epidemic_2017,kumaresan_fitting_2023,venkatramanan_using_2018}
\newcommand{\abmEpidemicPreparedness}[0]{woodul_agent-based_2023,marini_enhancing_2020}
\newcommand{\abmUrbanAreas}[0]{macal_chisim_2018,hackl_epidemic_2019}
\newcommand{\abmVaccinationHesitancy}[0]{martono_agent-based_2024,buttenheim_provider_2013}
\newcommand{\abmEpidemiologyOther}[0]{hunter_taxonomy_2017,miksch2019,estill2020,tracy2018}

\newcommand{\abmEpidemiology}[0]{\abmCOVID,\abmInfluenza,\abmMalaria,\abmHIV,\abmPublicHealthInterventions,\abmEpidemicPreparedness,\abmUrbanAreas,\abmVaccinationHesitancy,\abmEpidemiologyOther}

\newcommand{\abmSocialSciences}[0]{retzlaff2021,axelrod1981,arthur1994,axelrod1997,hegselmann2002,fischbach2021,schelling_segration_1971,epstein1996growing}
\newcommand{\abmFinanceAndEconomics}[0]{axtell2024,palmer_artificial_1994,farmer2009}
\newcommand{\abmAgriculture}[0]{huber2018,kremmydas_review_2018,yamashita_development_2018,maes_agent-based_2017,schreinemachers_agent-based_2011,barbuto_improving_2019,bert_agent_2011,coronese_agrilove_2023,beckers_modelling_2018}
\newcommand{\abmTransport}[0]{li2021,huang2022,nguyen2021,tzouras2023,kim_estimating_2022,cunha_development_2022,he_microscopic_2020,rindsfuser_agent-based_2007,wang_prediction_2021,tumer_distributed_2007}
\newcommand{\abmEcology}[0]{grimm2005,grimm2013,deangelis2005,zhang2020,grimm_ten_1999,deangelis_individual-based_2018,deangelis_individual-based_2014,judson_rise_1994,uchmanski_individual-based_1996,mclane_role_2011,karsai_bottom-up_2016,heckbert_agent-based_2010}
\newcommand{\abmEnergy}[0]{tian_agent-based_2020,robinson_determinants_2015,sachs_agent-based_2019,moglia_agent-based_2018,rai_agent-based_2016}
\newcommand{\abmMarketing}[0]{north_multiscale_2010,romero_two_2023,schramm_agent-based_2010,negahban2014,rand_agentbased_2021,garcia_validating_2007,rand_agent-based_2011}
\newcommand{\abmCrime}[0]{malleson_analysis_2012,zhu_agent-based_2021,malleson_crime_2010,groff_state_2019}
\newcommand{\abmAnimalBehavior}[0]{reynolds_flocks_1987}
\newcommand{\abmOther}[0]{boone2017,castro2020,stieler2022,heppenstall2021,macal_introductory_2014}

\newcommand{\abmAll}[0]{\abmBiology,\abmMedicine,\abmEpidemiology,\abmSocialSciences,\abmFinanceAndEconomics,\abmAgriculture,\abmTransport,\abmEcology,\abmAnimalBehavior,\abmMarketing,\abmEnergy,\abmCrime,\abmOther}

\newcommand{\mechanisticModels}[0]{baker_mechanistic_2018,craver_when_2006,curtis_mechanistic_1991,sedghi_taxonomy_2021,zhang_chapter_2010,wang_evolution_2023,muldbak_digital_2022}

\newcommand{\alphafold}[0]{jumper_highly_2021,yang_alphafold2_2023,varadi_alphafold_2022,senior_improved_2020}
\newcommand{\surrogateModels}[0]{angione_using_2022,cheng_review_2024,ganti_design_2020,garzon_machine_2022,cai_surrogate_2021,cozad_learning_2014,agarwal_machine-learning-based_2020,luo_review_2023,mai_machine_2021,weber_technical_2020,lu_efficient_2019,bocquet_surrogate_2023,zahura_training_2020,prina_machine_2024,zhu_building_2022}

\newcommand{\physicsInformed}[0]{batuwatta-gamage_physics-informed_2022,karniadakis_physics-informed_2021,sharma_review_2023,kashinath_physics-informed_2021,karimpouli_physics_2020,zhao_parameter_2022,nghiem_physics-informed_2023,pateras_taxonomic_2023,qian_lift_2020}
\newcommand{\ddmOverview}[0]{sarker_ai-based_2022,casalino_ai-driven_2021,noe_machine_2020,bernetti_data-driven_2020,noauthor_rise_2021,ourmazd_science_2020,baker_mechanistic_2018}

\newcommand{\bdmUsage}[0]{duswald_2024,demontigny_2021,\cognoLungInjury,duswald2024calibrationstochasticagentbasedneuron,hesam_2024,demontigny_2023,demetriades_interrogating_2022,gazeli_interrogating_2022,bdm-workshop,abubacar_neuronal_growth,jennings_cryo,jennings_cryo2,hiv_malawi}

\begin{abstract}

Agent-based simulation is a method for studying highly complex systems in disciplines as diverse as biology, medicine, sociology, economics, finance, and more. In this method, an agent represents a small part of the system, such as a human cell, and, through interactions among many agents, the system's behavior emerges, like the growth of cancer in the human body.
Real-world complex systems, such as a human organ, comprise billions of agents. However, the state-of-the-art agent-based simulation platform, BioDynaMo, does \emph{not} support the distributed processing of a simulation across servers. This limitation restricts the scalability of agent counts to the computational resources available within a single server, hindering users from conducting larger and more meaningful simulations.

In this paper, we 
introduce TeraAgent, a new distributed agent-based simulation engine.
To overcome BioDynaMo's scalability limitation, TeraAgent enables efficient exchange of agent information across servers by implementing two solutions: 1)~a tailored serialization mechanism that allows agents to be accessed and mutated directly at the receive buffer, and 2)~delta encoding that significantly reduces data transfer across servers by leveraging the iterative nature of agent-based simulations.
Built on our solutions, \ta{} 1) enables extreme-scale simulations with \ESMaxAgentsInTrillion{} agents ($\ESFactorImprovementSOTA\times$ that of \bdm{}) on \ESNumCPUs{} CPU cores, 2) reduces time-to-result leveraging parallelism across compute nodes, 3) improves interoperability with third-party tools, and 4) provides users with more hardware flexibility.
\ta{} serializes agents with a median speedup of $110\times$ ($37\times$ for deserialization) over the state-of-the-art ROOT I/O baseline, delivers $8\times$ the agent update rate per CPU core of Biocellion, and improves in-situ visualization performance by $39\times$.
To foster future research and enable simulation-based research in many fields, we will open-source \ta{}.

 \end{abstract}

\titlepgskip=-21pt
\maketitle

\section{Introduction}
Agent-based simulation is a bottom-up simulation method, where a complex system is modeled as a set of agents and {interactions across those agents, i.e., an agent-based model (ABM).} 
Many complex systems have been modeled and extensively studied as ABMs in various research fields~\cite{\abmAll}
(e.g., biology~\cite{\abmBiology}, medicine~\cite{\abmMedicine}, epidemiology~\cite{\abmEpidemiology}, political sciences~\cite{schelling_segration_1971, farmer2009}, complex social phenomena~\cite{epstein1996growing}, finance~\cite{\abmFinanceAndEconomics}, and animal behavior~\cite{reynolds_flocks_1987}).

Despite the vast diversity of simulated systems, building ABMs typically follows a common three-step structure.
First, the researcher defines the abstract term \emph{agent}. As an intuitive example, an agent can represent a bird in the complex system of a flock of birds~\cite{reynolds_flocks_1987}. 
In this definition, the agent corresponding to a bird maintains the attributes of its current position and velocity vector. 
Second, the researcher defines the agent's behaviors. 
For example, a bird avoids collision, aligns its velocity with its neighbors, and tries to stay close to other birds in the flock.
Third, the researcher defines the initial values of each agent's attributes, i.e., initial conditions. 
Once built, a simulation engine simulates the ABM for a number of iterations, or until a specific condition is met.

Each iteration simulates local interactions across agents (e.g., birds flying in close proximity to each other), resulting in emergent system behaviors (e.g., swarm dynamics of the flock of birds). This feature of ABMs enables distributed execution via spatial partitioning of agents, where data exchange across partitions happens at the bordering regions.

The future of ABM is exceptionally promising, with research actively advancing toward unprecedented simulation scales. There is strong scientific motivation to enable simulations that are much larger and more detailed. Biological systems provide compelling inspiration: for example, the human cortex contains approximately 16 billion neurons \cite{azevedo_equal_2009}, with each neuron potentially represented by several hundred agents \cite{zublerdouglas2009framework,zublerdissertation,torben-nielsen_context-aware_2014}. Recent studies underscore this ambition in practice. De Montigny et al. \cite{demontigny_2021} simulated glioma growth at cellular resolution and had to employ multiscale modeling and averaging techniques to make the computation feasible. Similarly, Duswald et al. \cite{duswald_2024} modeled vascular tumor growth up to $400~mm^{3}$ and reported that computational limitations prevented further scaling. These cases illustrate both the enormous potential and the active scientific demand to enable high-performance and scalable simulations of systems with hundreds of billions of agents.
Unfortunately, \emph{no} simulation platform provides such scalability. \bdm{}~\cite{breitwieser-bdm, breitwieser_biodynamo_2023} is the 
best-scaling state-of-the-art simulation platform (three orders of magnitude faster than Cortex3D and NetLogo and nearly an order of magnitude more efficient than Biocellion) and is already adopted in a wide range of disciplines~\cite{\bdmUsage}. However, \bdm{}'s scalability is limited by the computational, memory, and storage resources of a single server by design. 
In particular, researchers who use \bdm{} face
three critical restrictions:

\head{Limited simulation size} The potential to scale up simulations using more powerful hardware has its limits. Once the demand for memory and computing capacity exceeds the capabilities of high-end servers, further scaling becomes infeasible. Specifically, \bdm{}'s analysis~\cite{breitwieser_biodynamo_2023} reports that simulating more than \ESAgentsBDM{} agents on a single-server setup is impractical, even for high-end servers equipped with terabyte-sized main memory. This key constraint is particularly relevant since ABMs may involve hundreds of billions of agents.

\head{Inefficient interoperability with third-party software}
\bdm{}'s parallel computing is based purely on Open Multi-Processing (OpenMP) and thus it does \emph{not} efficiently interoperate with very relevant third-party software that is built on the standard Message Passing Interface (MPI) (e.g., ParaView~\cite{paraview}, particularly the in-situ visualization mode~\cite{paraview-insitu}, and OpenLB~\cite{krauseOpenLBOpenSource2021}).

\head{Inflexible allocation of hardware resources}
Even for simulations that would fit within the performance envelope of a high-end server, it may be more cost-effective to distribute the simulation across many lower-cost computational nodes.
Doing so requires the simulation platform to efficiently scale out, while \bdm{} is limited to a single server.

To overcome these three restrictions, we present \ta{}, a new \emph{distributed} simulation engine that efficiently scales out a single ABM simulation from a single computation node to \emph{many} computation nodes.
\ta{} achieves this by leveraging three key insights based on our domain-specific knowledge of ABM simulations: 
1)~ABM simulations are based on local interactions of agents and thus can be spatially partitioned, 2)~the iterative nature of ABM simulations causes the agent attributes to change gradually across iterations, and 3)~relevant third-party software predominantly adopts an MPI-based implementation. 

First, \ta{} divides the simulation space
of an ABM into multiple smaller partitions
based on the spatial distribution of the agents in the ABM, and distributes the simulation across multiple computation nodes. Such distributed computing leverages the first key insight, such that most of an agent's interactions are local to the ABM partition where the agent resides. 
Such distributed execution requires the exchange of agent information across partitions 1)~to obtain the local environment for an agent update and 2)~to migrate an agent across partitions when necessary, e.g., when an agent's position changes during the simulation such that it crosses a partition boundary.
To communicate data across partitions, \ta{} implements a serialization mechanism that packs agents as objects into a contiguous buffer while maintaining full mutability of their data structures.

Second, \ta{} reduces the performance overheads of communication across nodes by leveraging the second key insight.
To do so, \ta{} implements
a delta encoding-based compression scheme (similar to prior work~\cite{pekhimenko_base-delta-immediate_2012}) that effectively reduces data transfers from absolute values of agent features to the deltas of those values across iterations.

Third, \ta{} implements a hybrid interface based on MPI and OpenMP. To leverage the insight that relevant third-party software adopts MPI as the interface, \ta{} adopts MPI as an ABM partition's interface to other partitions and third-party software. Thus, ABM partitions and third-party software natively use the same interface.
\ta{} adopts \bdm{}'s OpenMP-based parallelism to simulate the local interactions across agents within an ABM partition to leverage the performance benefits of shared-memory-based parallelism. \ta{}'s hybrid implementation enables high-performance and scalable agent-based simulation.

We rigorously 1)~evaluate \ta{}'s functionality; 2)~compare it against two state-of-the-art platforms: \bdm{} and Biocellion in terms of scalability and interoperability; and 3)~break down \ta{}'s improvements into serialization
and delta encoding. 
Our evaluation results show that \ta{} simulates \result{\ESFactorImprovementSOTA}$\times$ more agents than \bdm{}, achieves an \result{8}$\times$ higher agent update rate per CPU core than Biocellion~\cite{kang2014biocellion}, and improves the visualization performance by \result{39}$\times$.
The median serialization performance increases by \result{110}$\times$ (deserialization by \result{37}$\times$), and delta encoding reduces the required data transfer by up to \result{3.5}$\times$.
\ta{} 1)~enables extreme-scale simulations with \ESMaxAgentsInTrillion{} agents (i.e., an \ESFactorImprovementSOTA$\times$ improvement over \bdm{} and an \ESFactorImprovementLiterature$\times$ improvement over the largest agent count previously reported in the literature~\cite{parker_2007}) on \ESNumCPUs{} CPU cores, 2)~reduces time-to-result by adding compute nodes, 3)~improves interoperability with third-party tools in terms of performance, and 4)~provides users with greater hardware flexibility.

This paper makes the following major contributions.

\begin{itemize}
  \item We show that it is possible to make scale-out agent-based simulations practical by alleviating their data movement overheads across nodes via tailored serialization and delta encoding-based compression.
\item We present a novel distributed simulation engine called \ta{} capable of simulating \ESMaxAgentsExact{} agents and scaling
to \ESNumCPUs{} CPU cores.
        This number of CPU cores represents the maximum we were permitted to use on the Dutch National Supercomputer.

	\item We present a novel serialization method for the agent-based use case, which significantly reduces the time spent on packing and unpacking agents during agent migrations and aura exchanges.
	      We observe a median speedup of \result{110}$\times$ for serialization and
	        \result{37}$\times$ for deserialization.
\item We extend our serialization method to support delta encoding to reduce the amount of data that has to be transferred between servers, reducing the message size by up to \result{3.5}$\times$.

\end{itemize}

\section{Background and Motivation}

\label{sec:background_and_motivation}

We provide a brief overview of agent-based models (ABMs), ABM simulations, parallel processing based on OpenMP, and distributed execution based on MPI. This section focuses on concepts relevant to scaling agent-based simulations instead of providing a broad background of these topics.

\subsection{Agent-based simulation}

Agent-based simulation is a bottom-up simulation methodology where a complex system is modeled as a set of simple agents and their interactions with other agents and the environment. 

\head{Agent}
An agent is an abstract representation of an entity in terms of its state, behavior, and interactions. The entity is an autonomous individual in a complex system, e.g., a biological cell in a tissue, a bird in a flock of birds, an investor in a stock market, or a patient in a human population.

\head{Agent-based model}
An agent-based model (ABM) specifies the set of agents in a system, their state variables, behavioral rules, interactions, and initial states. Two key concepts of ABMs are \emph{local interactions} and \emph{emergent system behavior}.

\head{Local interactions}
An agent in an ABM only interacts with its local environment and does \emph{not} possess information about agents far away. For example, a bird in a flock of birds \emph{only} observes the birds in its close proximity and adjusts its location and velocity vector accordingly \emph{without} any direct interactions with other birds in far locations.

\head{Emergent system behavior} 
Emergent behavior refers to complex patterns and phenomena that arise from the interactions of individual agents within a system. In agent-based models, this behavior is significant as it highlights how collective actions lead to outcomes that cannot be predicted by examining the agents in isolation. For instance, when agents interact with one another and their environment, they can create intricate dynamics at a macro level, which differ from the simple, individual behaviors observed at the micro level. This concept embodies the principle that ``more is different''~\cite{anderson1972}, showing that the whole can be greater than the sum of its parts.
Emergence can manifest in various forms, much like the diverse range of agents and behaviors resulting from different use cases. 
Examples of emergent behavior include flocking dynamics in birds \cite{reynolds_flocks_1987}, pandemic hospitalization patterns \cite{hesam_2024}, traffic
jams \cite{nguyen2021}, wealth distribution \cite{epstein1996growing}, and many more.

\head{Agent-based simulation}
An agent-based simulation executes an agent-based model in an iterative way, where each agent's state is updated based on the agent's behavior and local interactions. Agents can move, interact, create or destroy other agents, or modify shared environmental state across iterations. The computational cost of a simulation depends on the agent count and the complexity of agent behavior and interactions.

\head{Distributed execution of agent-based simulations}
By modeling a complex system as a set of agents and their local interactions, ABM simulations are inherently parallelizable via spatial partitioning of agents. In principle, such parallel execution of ABM simulations can be widely distributed across many computation nodes, where most of the interactions take place with high spatial locality within a computation node and relatively fewer interactions in bordering regions need to be simulated across partitions.

\head{Broader scope of agent-based simulations}
Agent-based simulations are used across a wide range of scientific and engineering domains, including biology~\cite{\abmBiology}, medicine~\cite{\abmMedicine}, epidemiology~\cite{\abmEpidemiology}, political sciences~\cite{schelling_segration_1971, farmer2009}, complex social phenomena~\cite{epstein1996growing}, finance~\cite{\abmFinanceAndEconomics}, and animal behaviors~\cite{reynolds_flocks_1987}. In many of these domains, simulations are used to explore scenarios that are difficult or impractical to study experimentally, to test hypotheses, or to analyze the emergent behavior of complex systems under varying conditions.
Scaling ABMs and reducing their infrastructure costs would enable simulating more detailed and more complex systems in these fields.
Such scaled and low-cost simulations would accelerate scientific progress by enabling researchers to evaluate their ideas quickly and to gain deeper insights into the emergent behaviors of complex systems.

\subsection{Parallel Execution using OpenMP and MPI}

In the scope of this paper, we consider parallelism in two different domains: shared-memory-based parallel execution built on OpenMP, and distributed execution built on MPI.

\head{Shared-memory-based parallel execution, built on OpenMP}
Multi-threaded execution allows a single program to execute multiple threads of control concurrently within one process. Threads share the same address space and can communicate through shared variables. 
In a shared-memory system, multiple threads or processing cores can directly access the same physical memory. This programming model enables low-latency communication through memory loads and stores, but it also limits the scalability of parallelism by the resources available in a computation node.
OpenMP is a widely used programming model for shared-memory parallelism that provides compiler directives, runtime routines, and environment variables for expressing parallel regions, work sharing, and synchronization. OpenMP enables incremental parallelization of existing code and is commonly used to exploit multicore CPUs. 

\head{Distributed execution, built on MPI}
Scale-up refers to improving performance by using more resources within a single system, such as additional cores or larger memory capacity. Scale-out refers to increasing performance by distributing computation across multiple systems connected by a network. While scale-up benefits from low-latency shared memory, it is constrained by hardware limits. Scale-out systems offer greater aggregate resources but require explicit management of communication, data distribution, and synchronization.
The Message Passing Interface (MPI) is a standardized programming model for distributed-memory parallel computing. MPI programs consist of multiple processes, each with a private address space, that communicate explicitly by sending and receiving messages. MPI provides point-to-point communication, collective operations, and synchronization primitives, and is designed to be portable and efficient across a wide range of hardware interconnects. MPI is widely used in high-performance computing applications that require scalable execution across many nodes.

\subsection{Motivation}

Many scientific domains rely on agent-based simulations to study systems with large populations, fine spatial resolution, or long simulation horizons. As models grow in scale and fidelity, the computational and memory demands of these simulations increase correspondingly. In practice, simulations with billions of agents often exceed the capabilities of single-node systems, motivating the use of parallel and distributed computing resources.
Despite the natural parallelism present in agent-based simulations, achieving scalable performance remains challenging. Many existing frameworks primarily target shared-memory systems and rely on multi-threading to exploit parallelism, which limits scalability to a single node. Distributed execution introduces additional challenges, including spatial partitioning, communication overhead, data consistency, and load imbalance due to dynamic agent behavior. Addressing these challenges in a way that preserves usability, performance, and correctness remains an open problem for large-scale agent-based simulation systems.

\section{TeraAgent}
\label{sec:design-overview}

We present \ta{}, a new distributed agent-based simulation engine. \ta{} 1)~partitions an agent-based simulation based on the spatial locality of agent interactions and distributes across many computation nodes, 2)~implements efficient and scalable communication across those partitions via its serialization and compression schemes that leverage gradual state changes across the simulation's iterations, and 3)~provides the distributed ABM with a native integration with relevant third-party software via MPI, improving the efficiency of its interoperability with such software. This section provides an overview of \ta{} (\secref{sec:design:distribution}), presents \ta{}'s serialization mechanism (\secref{sec:design:serialization}) and compression scheme (\secref{sec:design:delta-encoding}), explains \ta{}'s implementation details (\secref{sec:design:implementation}), and discusses \ta{}'s other contributions to the state-of-the-art (\secref{sec:design:smpbdm}).

\subsection{\ta{} Overview}
\label{sec:design:distribution}

\begin{figure}[bth]
\includegraphics[width=.9\linewidth]{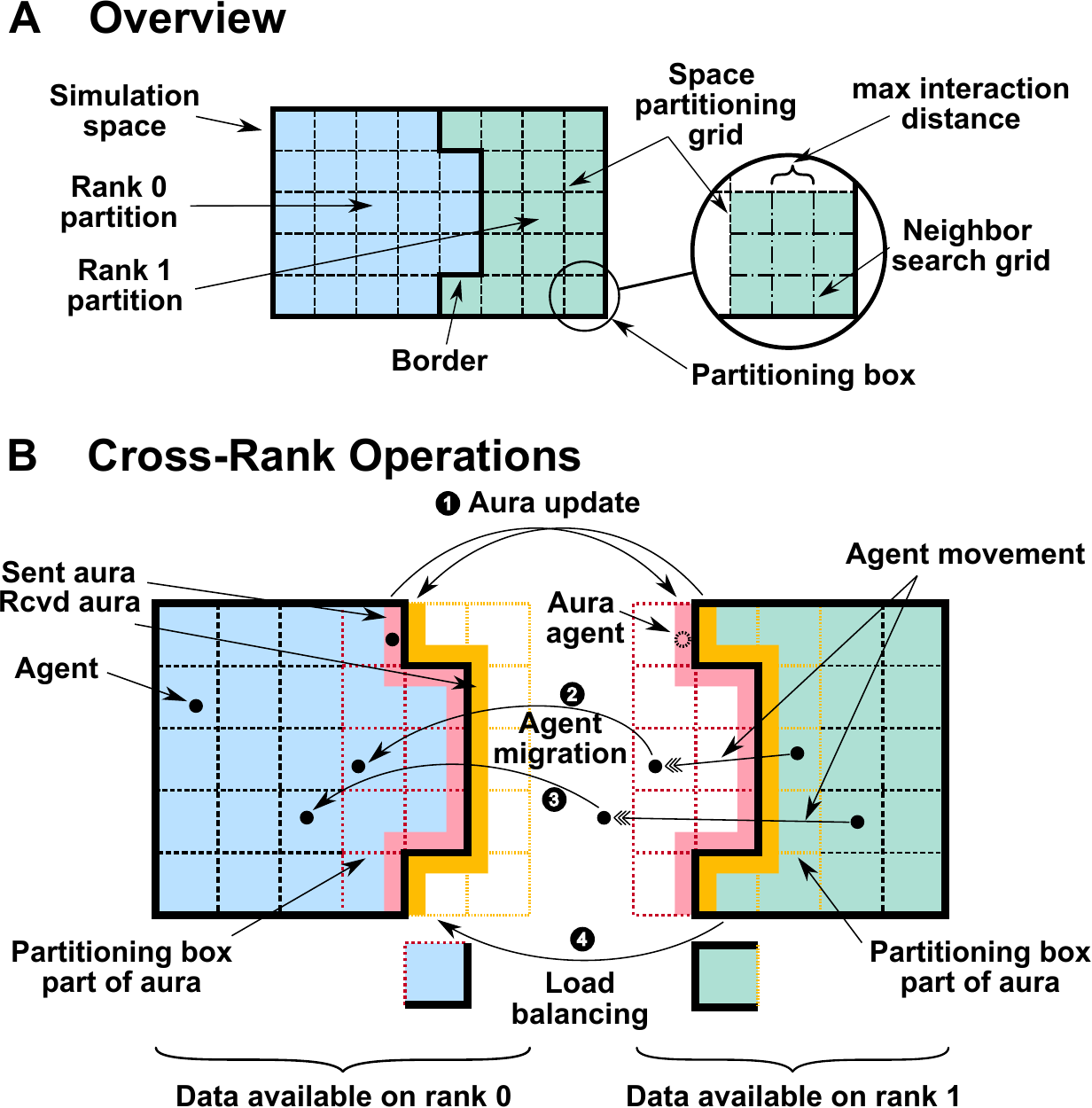}
	\caption{Overview of \ta{}'s distributed execution, simplified to two dimensions and two ranks. (A)~Space partitioning and the neighbor search grid. (B)~The four cross-rank operations: 1)~aura update, 2)~agent migration where the moving agent lands within the aura region, 3)~agent migration where the moving agents land outside any locally known aura region, and 4)~load balancing.}
\label{fig:design:overview}
\end{figure}

\figref{fig:design:overview} shows a general overview of \ta{}'s distributed execution.
\figref{fig:design:overview}A demonstrates the partitioning of a 2D ABM and its distribution across two processes (i.e., MPI ranks~\cite{mpi}) that do \emph{not} share their memory spaces. The figure is simplified to two dimensions for clarity; the same mechanism applies to 3D simulations.
\ta{} first partitions the simulation space into smaller mutually exclusive volumes by applying a 2D grid called the \emph{space-partitioning grid} (shown with dashed lines in \figref{fig:design:overview}A on the left-hand side). 
In this grid, each partition (shown with different colors and a bold border in \figref{fig:design:overview}A) comprises multiple \emph{partitioning boxes} and is exclusively assigned to an MPI rank. 
Within a partitioning box, a finer-grained grid called the \emph{neighbor search grid} is applied. Each cell of the neighbor search grid is as wide as the maximum interaction distance, such that an agent can interact \emph{only} with the agents in the same or adjacent cells in the neighbor search grid. This fine-grained simulation space partitioning allows agents to efficiently find their neighbor agents and simulate their interactions.
In this example, a partitioning box is split into 9 cells by a $3\times3$ neighbor search grid.

\ta{} implements parallel execution of agent interactions using OpenMP within a partitioning rank, and MPI across partitioning ranks to achieve both the high performance of a shared-memory-based implementation within a computation node and efficient scalability across computation nodes.

To simulate agent interactions at the bordering regions, \ta{} implements a number of operations across rank partitions. 
\figref{fig:design:overview}B shows the major operations across two rank partitions (colored in blue on the left-hand side and green on the right-hand side). Each dot in \figref{fig:design:overview}B represents an agent, which may or may not be close to the bordering region of the rank partition. Although two agents in the bordering regions of two different rank partitions are simulated on different computation nodes, they can be within each other's interaction distance in the simulation space and their interactions should be simulated. To efficiently simulate such interactions, \ta{} creates a shadow of such agents in the opposite rank partitions. This is done by extending a rank partition with a shadow copy of the bordering region of the opposite rank partition, called an \emph{aura} (also called a halo or ghost in the literature). The shadow copies of the agents in an aura are accordingly called \emph{aura agents}. \ta{} performs four major operations to maintain aura agents and efficiently and correctly simulate agent interactions in bordering regions: 1)~aura update, 2)~agent migration within the bordering region, 3)~agent migration from outside the bordering region, and 4)~load balancing.

\head{Aura update (\circled{1})} Aura agent states should be synchronized across rank partitions at the end of an iteration for correctness. Such operations are called \emph{aura updates}. 

\head{Agent migrations (\circled{2} and \circled{3})} An agent may change its position during a simulation iteration and it can move across rank partitions. In such a case, \ta{} migrates the agent to its new rank, i.e., \emph{agent migration}. We consider two possible cases separately. First, the destination of agent migration can be in the aura of the agent's current rank partition (\circled{2}).
Here, the new rank is already known, allowing the agent to be transferred directly to its new destination.
Second, the agent's new location falls outside all local auras (\circled{3}). In such a case, \ta{} performs a collective lookup stage to determine the agent's appropriate new rank before migrating the agent. 

\head{Load balancing (\circled{4})} Agent migrations during the simulation can lead to load imbalances across MPI ranks, reducing resource utilization. To provide high resource utilization and improve load balancing, \ta{} adjusts rank partitions online and performs agent migrations accordingly.

To efficiently perform data movement across MPI ranks due to aura updates, agent migrations, and load balancing, \ta{} implements a serialization mechanism and a compression scheme.

\subsection{Serialization}
\label{sec:design:serialization}

\ta{} needs a serialization mechanism to coalesce data objects and transfer them across MPI ranks in batches to reduce the cost of inter-rank communication per bit.
\bdm{}~\cite{breitwieser-bdm, breitwieser_biodynamo_2023} and the field of high-energy physics use ROOT IO~\cite{brun_root_1997,root-files} for serialization, as it provides higher performance than other alternatives \cite{blomer_quantitative_2018} (e.g., Protobuf \cite{protobuf}, HDF5 \cite{hdf5}, Parquet \cite{parquet}, and Avro \cite{avro}).
However, we observe that agent serialization using ROOT IO creates major performance bottlenecks, and we propose a new serialization mechanism called \ta{}-IO. We explain four reasons for ROOT IO's performance bottleneck in the ABM use case and how \ta{}-IO addresses them.

First, ROOT implements pointer deduplication, which is not needed for \ta{}. ROOT IO keeps track of already seen pointers during serialization so that ROOT can skip over repeated occurrences of the same pointer and ensure that upon deserialization all pointers point to the same instance. However, \ta{} does \emph{not} allow multiple agents to have pointers to the same object, and thus does \emph{not} need pointer deduplication. Therefore, we safely eliminate pointer deduplication and its overheads. 
To support agents in mutating the state of their neighbors, \bdm{} provides a smart-pointer class called \texttt{AgentPointer}.
An \texttt{AgentPointer} object stores the unique agent identifier of the referenced agent and obtains the raw pointer from a map stored in the \texttt{ResourceManager}.
Therefore, \texttt{AgentPointer} objects are serialized as unique agent identifiers. \ta{} supports \emph{only} \texttt{const} \texttt{AgentPointer}s, which avoids cross-rank merges.

Second, we observe that the long latency of deserialization creates a performance bottleneck. 
To perform deserialization faster, \ta{} adopts key ideas from Google's FlatBuffers serialization library~\cite{flatbuffers}, most notably the ability to 1)~access serialized data directly \emph{without unpacking} and 2)~support mutable object attributes. 
While FlatBuffers supports some mutability, it remains limited in scope.
\ta{} addresses this limitation by enabling full mutability of object attributes required by the simulation engine, thereby removing the restrictions present in FlatBuffers.

Third, ROOT IO implements a conversion across big and little endianness. However, distributed execution across computation nodes that implement different endianness is out of \ta{}'s scope, and thus this conversion is \emph{not} needed. Therefore, \ta{}-IO does \emph{not} implement such conversion and avoids its overheads.

Fourth, ROOT IO implements a schema evolution mechanism to ensure data integrity across large time windows (e.g., decades). However, \ta{}'s data exchange across ranks inherently does \emph{not} require such strong guarantees as the data is \emph{not} needed after the simulation or even after one or a few iterations. 
Therefore, \ta{} skips schema evolution to reduce overheads.

\head{\ta{}-IO} \label{sec:design:bdmio}
Based on the four observations and the design principle that ``What you don't use you don't pay for''~\cite{stroustrup1994design}, we design \ta{}-IO, a low-overhead serialization mechanism crafted for the needs of distributed agent-based simulations.

\begin{figure}[bth]
\includegraphics[width=.8\linewidth]{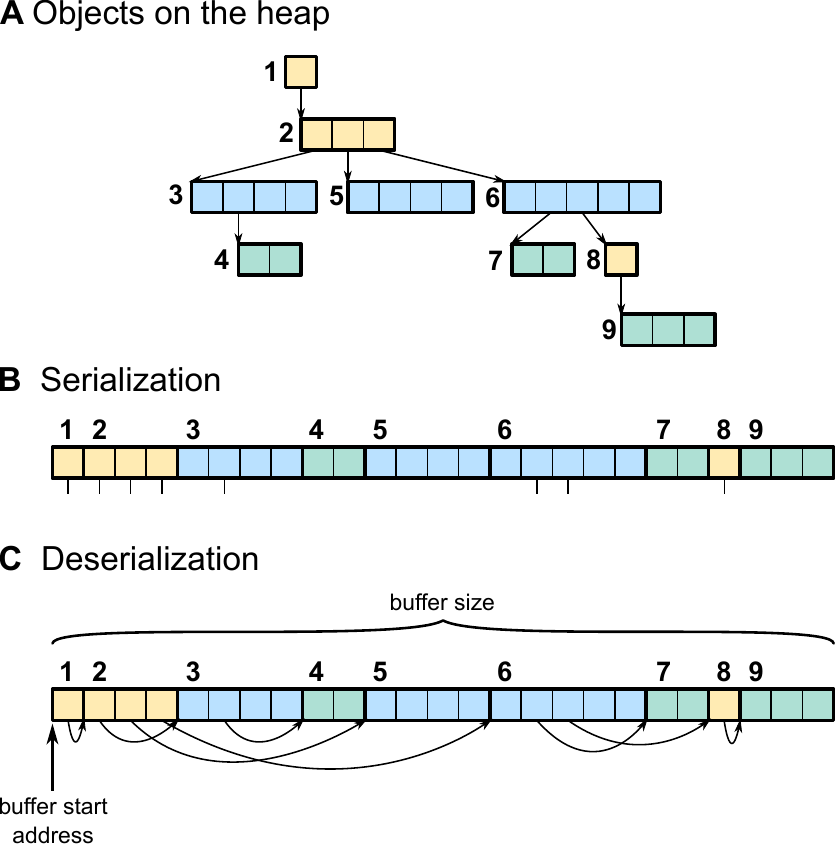}
	\caption{The \ta{}-IO serialization mechanism. Numbered nodes in~(A) are contiguous memory blocks and edges are pointers; the arcs in~(C) are pointer fields repointed in place, with no further allocation.}
	\label{fig:design:serialization}
\end{figure}

\figref{fig:design:serialization}A shows an abstract representation of the objects in memory maintained in a tree structure.
The nodes of the tree are formed by contiguous memory blocks that were allocated on the heap.

\head{Serialization}
\label{sec:design:bdmio-serialization}
This structure can be serialized by a pre-order (depth-first) tree traversal (\figref{fig:design:serialization}B).
Each tree node (i.e., memory block) is copied to the serialized buffer.
Fields inside each memory block that point to other memory blocks are labeled as such, but point to the invalid address \texttt{0x1}.
The virtual table pointer of polymorphic classes is replaced by a
  unique identifier of the most derived class.
This step is necessary because we cannot guarantee that the virtual table
  pointers will be the same on all ranks.

\head{Deserialization}
From the communication subsystem, the deserialization implementation receives a buffer with a starting address, length, and type of the root object (\figref{fig:design:serialization}C).
Deserialization consists of four steps that can be performed in a single
  traversal of the buffer.
First, we traverse the tree from the beginning of the buffer and restore the
  virtual table pointer for polymorphic objects.
The offsets of the memory blocks are determined by their type. 
Second, if a pointer is encountered, we set it to the next memory block in the
  buffer.
Third, as we traverse, we count the number of memory blocks, which will be
  required for memory deallocation (see next paragraph).
Fourth, we reinterpret the buffer's starting address as a pointer to the root
  object and return it to the caller.
No other memory reads or writes take place.
Furthermore, there are no calls to allocate memory besides the single
  contiguous receive buffer that holds the data.

\head{Mutability}
By returning the root object pointer to the higher-level code, we create the
  illusion that all contained memory blocks have been allocated separately on the
  heap.
Therefore, higher-level code is not aware that these objects were deserialized
  using \ta{}-IO and can change them in any way.
This includes setting the values of attributes, but extends to, for example, adding
elements to containers, such as \texttt{std::vector}, even if there is not sufficient space in the buffer.
In this scenario, the vector implementation notices that the capacity is reached,
  allocates a new memory block on the heap (separately from the buffer) and
  deallocates the obsolete memory block inside the deserialized buffer.

\head{Deallocation}
As we have seen in the vector example above, higher-level code will at some
  point try to deallocate memory blocks by calling \texttt{delete}.
These \texttt{delete} calls, however, would crash the memory allocator, due to the
  missing corresponding \texttt{new} call.
To maintain the illusion, we intercept all calls to \texttt{delete} and filter those
  that fall into the memory range of the deserialized buffer.
If the number of expected \texttt{delete} calls (determined during deserialization)
  matches the intercepted delete calls, we deallocate the whole buffer and remove
  the filter rule.

\ta{} does \emph{not} leak memory, because all memory blocks in the deserialized buffers are freed, for the following two reasons.
First, the aura region is completely rebuilt in each iteration, which means
  that the previous aura information is completely destroyed.
Second, for agent migrations and load balancing data, we rely on the periodic
  agent sorting mechanism inherited from \bdm{}.
Agent sorting changes the memory location of agents to improve the cache hit
  rate.
During this process all agents are copied to a new location and the old ones
  are deallocated.
Our evaluation shows that the memory consumption does not 
  increase by using the \ta{}-IO serialization mechanism.

\subsection{Data Compression}
\label{sec:design:delta-encoding}

\begin{figure}[tbh]
\includegraphics[width=\linewidth]{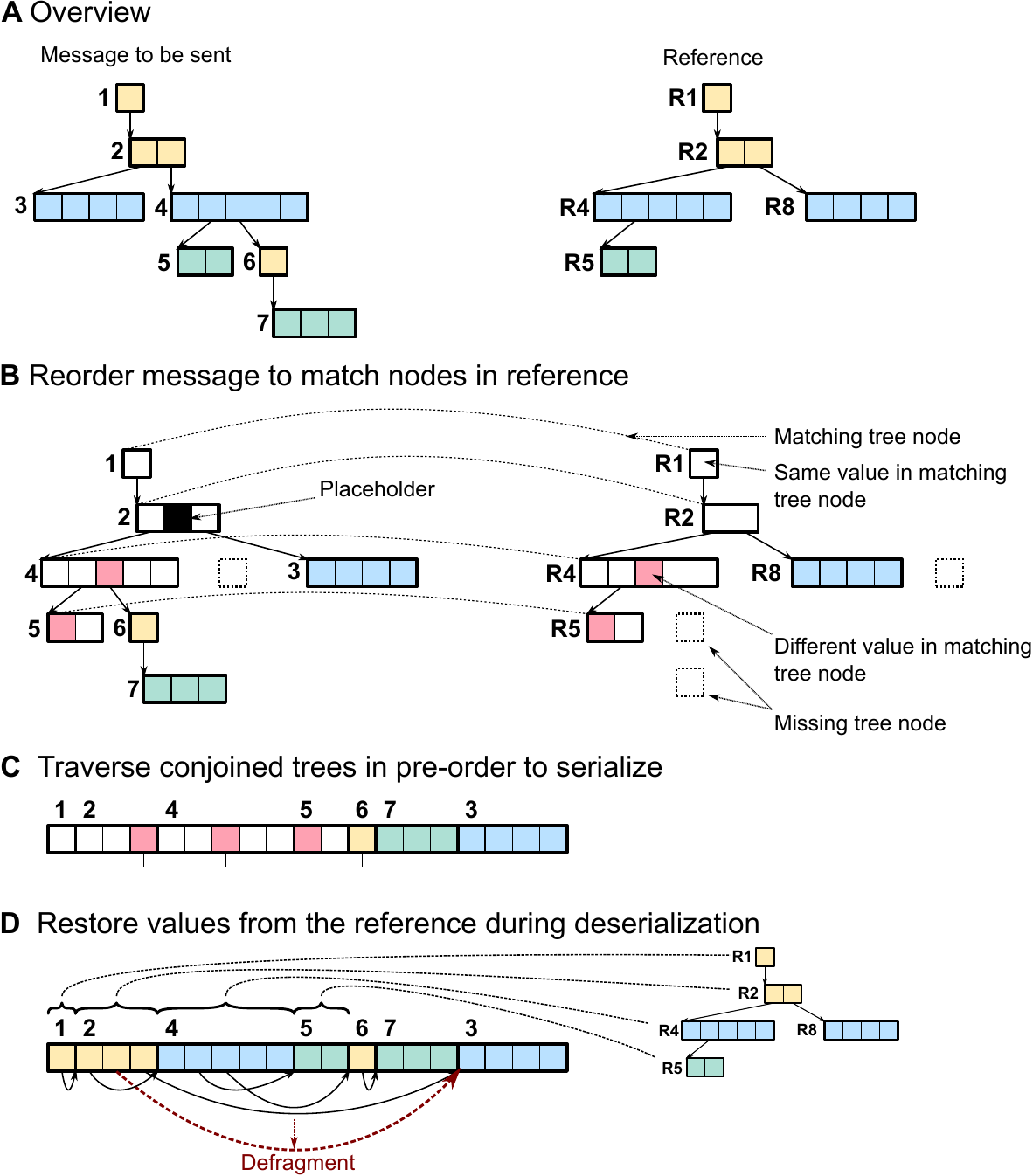}
	\caption{\ta{}'s delta encoding scheme. The reference is identical on the sender and the receiver. $\mathrm{R}n$ corresponds to node $n$ in the message.}
	\label{fig:design:delta-compression}
\end{figure}

ABMs are iteratively simulated and agent states change \emph{only} gradually across consecutive iterations. 
We leverage this observation to reduce the amount of data that must be transferred between MPI ranks to update the aura region by using delta encoding, similar to prior work~\cite{pekhimenko_base-delta-immediate_2012}.
\figref{fig:design:delta-compression} depicts our delta-encoding-based compression scheme, applied to the \ta{}-IO serialization mechanism.
\figref{fig:design:delta-compression}A shows the message and a reference.
The reference is identical on the sender and receiver ranks.
The sender calculates the difference between the message and the reference, compresses this difference, and sends it. The receiver uncompresses the received message and restores the original message by applying the difference to its copy of the reference.
With additions and removals of agents, the reference can change drastically as the simulation progresses.
To keep the transmitted deltas small, the sender and the receiver update their references at regular intervals (by default every 50 iterations) and reorder the message to match the reference.

\figref{fig:design:delta-compression}B shows the message reordering to match the message's structure to the reference. The sender reorders the message at the agent-pointer level.
Agents in the message are moved to their corresponding places in the reference if they already exist in the reference. 
Agents that exist in the reference and not in the message are indicated by an empty placeholder, i.e., a null pointer in our implementation.
Lastly, agents that exist in the message and not in the reference are appended
  at the very end.
Reordering agents does not affect the correctness of a \ta{} simulation, because the order is not observed by the model, for the following two reasons.
First, migrated and rebalanced agents (i.e., agents owned by the rank) are integrated into \bdm{}'s single-node engine and follow the execution order depending on the modeler's discretization choices (see execution modes in \cite{breitwieserDesignAnalysisExtremeScale2024}).
Second, aura agents are currently read-only and are accessed by their location in space, rather than their position in the received message.

\figref{fig:design:delta-compression}C shows how the message is serialized with a traversal of two conjoined trees.
\ta{} writes the difference to the result buffer if there is a match.

\figref{fig:design:delta-compression}D shows deserialization and restoring original values using the reference.
\ta{} defragments the message by removing any placeholders that were
  inserted during the matching stage and passes the restored message to higher-level code.

By using this mechanism, \ta{} reduces the data movement bottleneck and increases the efficiency 
  of distributed execution.
  
\subsection{Implementation Details}
\label{sec:design:implementation}

\subsubsection{Partitioning Grid}
One important data structure in the distributed simulation engine is the
  partitioning grid, which is responsible for the domain decomposition of the
  simulation space as shown in Figure~\ref{fig:design:overview}.
We use Sierra Toolkit (STK) \cite{sierra-toolkit}, because it is (i)
  an established tool maintained as part of Trilinos \cite{trilinos-website},
  (ii) well integrated with the load balancing framework Zoltan2 \cite{zoltan2},
  and (iii) able to export the grid in exodus format~\cite{exodus},
  which our main visualization framework ParaView \cite{paraview} can read.
However, since STK is a generic mesh library, capable of much more than processing rectilinear grids, this functionality comes with a performance penalty in terms of compute and required memory.
Furthermore, although STK uses Kokkos \cite{kokkos} for shared memory
  parallelism, we observe that if \ta{} is executed in MPI-hybrid mode
  (MPI/OpenMP), most threads are idle during balancing and mesh modification
  calls into STK.
As an alternative, we also considered the parallel grid library
  \cite{honkonen_parallel_2013}, but experienced difficulties generating very
  large grid sizes.

To reduce the memory and compute footprint of STK's partitioning grid, we
  introduce a parameter to make the partitioning box length a multiple of the
  neighbor search grid (see Figure~\ref{fig:design:overview}).
The higher this factor, the less memory and compute resources are
  needed for the space partitioning, but at the cost of a coarser granularity at
  which load balancing decisions can be made.

\subsubsection{Serialization}
The \ta{}-IO mechanism (\secref{sec:design:bdmio}) requires that
  (de)serialization methods (also known as \ta{}-IO functions) are available for
  all types in the message.
\ta{} provides the necessary implementation for its internal classes, selected
  standard template library (STL) classes, and important groups of types (e.g.,
  polymorphic types, pointers, and plain arrays).
Due to the design decision to opt for a very lightweight deserialization
  method, the implementation of the (de)serialization methods depends on the
  internal attributes of each type (i.e., its concrete implementation), rather than on its public
  interface. 
This design decision might increase the maintenance effort.
However, the \ta{}-IO methods follow a regular structure and can be generated
  automatically in most cases.
This is especially true for all user-defined classes in this \ThesisPaper{chapter}{paper}.
Although the (de)serialization methods are currently written by hand, they can
easily be integrated into \bdm{}'s existing code generation stages, implemented at compile time and at runtime using the
C++ just-in-time compiler cling \cite{vasilev_cling_2012}.

Custom \\ta{}-IO functions are necessary if classes contain pointer attributes
  whose memory they do not own.
An example could be an array-based stack implementation with an attribute that
  points to the current top of the stack.
To this end, \ta{} allows for custom implementations of the \ta{}-IO functions that
  will replace the automatically generated ones.

To intercept delete calls, we override all global C++ \texttt{delete} operators and
  insert a call to the delete filter, which returns whether the \texttt{delete} should be
  filtered out or executed.
This design choice complicates combining \ta{} with performance or
  correctness tools (e.g., Valgrind) that use \texttt{LD\_PRELOAD} to inject
  their own implementation of \texttt{new} and \texttt{delete} and prevent the \ta{} ones from
  being executed. We addressed this issue by patching Valgrind.

\subsubsection{Communication}

For the majority of data transfers, we use non-blocking point-to-point
  communication (\texttt{MPI\_Isend}, \texttt{MPI\_Irecv}, and
  \texttt{MPI\_Probe}).
This choice allows for overlapping communication with computation to hide
  latency.
For regular communication patterns that occur between neighbors, we issue
  speculative receive requests right after the previous transfer has finished, to
  avoid delays caused by late receivers.
If the neighbors of a rank change due to load balancing, obsolete speculative
  receives are cancelled.
We transmit large messages in smaller batches to reduce the memory
  footprint of serialization and compression.

\subsubsection{Distributed Initialization}

Although the distributed simulation engine has the capability to migrate agents
  to any other MPI rank (\secref{sec:design:distribution}), we avoid
  a costly mass migration of agents during the initialization stage by creating
  agents on their authoritative rank whenever possible.
This is straightforward for regular geometric shapes (e.g., agents created on a
  surface defined by a function), but we also address agent populations that are
  created using a uniform random distribution within a specific space.
Each rank determines the fraction of the target space that overlaps with its
  authoritative volume and adjusts the agent count and bounds for that fraction
  accordingly, provided that every rank receives a sufficiently high agent count.

\subsubsection{Load Balancing}

Load balancing has two goals.
{First, partition the simulation space in a way that simulating one iteration
  takes the same amount of time on all ranks.
Second, the partitioning should minimize} the distributed overheads, e.g., the
  number of aura agents that must be exchanged.
\ta{} provides two classes of load balancing methods to achieve these goals:
  global and diffusive.

The global balancing method is based on STK~\cite{sierra-toolkit} and Zoltan2~\cite{zoltan2}.
We provide their functionality for \ta{} users and choose the recursive
  coordinate bisection (RCB) algorithm as default.
We apply a weight field on the partitioning grid and set the weight of each
  partitioning box based on the number of agents contained and scale it by the
  runtime of the last iteration.
Zoltan2 partitions the space such that the total weight of the partitioning
  boxes owned by each rank is uniform across ranks.
Doing so can lead to a new partitioning, substantially different from the previous one, causing mass migrations of agents to their new authoritative rank.

The diffusive approach migrates partition boxes from relatively slower ranks to their faster neighbor ranks.

\subsection{Other Contributions}
\label{sec:design:smpbdm}
This section describes the necessary modifications and improvements to the
  existing OpenMP-parallelized \bdm{} version to enable distributed execution.

\head{Parallelization Modes}
By building upon and extending the shared-memory capabilities, we provide two
  distributed execution modes: MPI-hybrid (MPI/OpenMP), and MPI-only.

For the MPI-hybrid mode, we launch one MPI rank for each NUMA domain on a
  compute node, while for MPI-only we launch one rank for each CPU core.
On modern hardware with ever-increasing CPU core counts, the MPI-hybrid
  mode reduces the number of ranks by almost two orders of
  magnitude.
We therefore expect the MPI-hybrid mode to be more efficient, which we confirm in \secref{sec:de:eval:bdm}.
The MPI-only mode, on the other hand, provides benefits when interfacing with
  third-party applications that are only parallelized using MPI.
In the MPI-hybrid mode, these applications would leave all but one thread per
  rank idle.
Switching between parallelization modes does not require recompilation of
  \ta{}.

\head{Unique Agent Identifiers}
\label{sec:design:agent-identifiers}
\bdm{} uses unique agent identifiers to address agents, since the actual memory
  location might change due to agent sorting.
Agent sorting improves the performance by reordering agents in a way that
  agents that are close in 3D space are also close in memory (see
  \cite{breitwieser_biodynamo_2023}).
The identifier comprises two fields, $\langle \mathit{index}, \mathit{reuse\_counter}
	  \rangle$, and has the following invariant: at any point in time, there is only
  one (active) agent in the simulation with a given index.
If an agent is removed from the simulation, this index will be reused, but to
  satisfy uniqueness the \texttt{reuse\_counter} is incremented.
This design allows the construction of a vector-based unordered map, where the
  first part of the identifier (the index) is used to index the vector.
This map allows for lock-less additions and removals to distinct map elements.

However, this solution does not work without changes for distributed executions
  with agent migrations and aura updates, because the invariant that the $\mathit{index}$
  values of the identifiers are almost contiguous no longer holds.
This would waste a considerable amount of memory in the vector-based unordered
  map.

Therefore, we rename the existing identifier to ``local identifier'' and
  also introduce a global identifier $\langle rank, counter \rangle$.
Rank is set to the rank where the agent was created, and ``counter'' is a
  strictly increasing number.
The global identifier of an agent is constant, but the agent might have various
  different local identifiers during the whole simulation.

We implement this change in a minimally invasive way.
The translation between local and global identifier happens only during
  serialization if the agent is transferred to another rank, or written to disk
  as part of a backup or checkpoint.
Global identifiers are only generated on demand.
If there are no backups or checkpoints and the agent stays on the same rank
  throughout the simulation, the agent will only have a local identifier.

\head{Incremental Updates to the Neighbor Search Grid (NSG)}
\ifthesis
Chapter~\ref{ch:engine} showed that our
  optimized uniform grid implementation performed best for the benchmarked
  simulations.
\else
Breitwieser \etal{} \cite{breitwieser_biodynamo_2023} showed that their
  optimized uniform grid implementation performed best for the benchmarked
  simulations.
\fi
Updates to the NSG required a complete rebuild, which had been adequate so far.
The distributed simulation engine, however, relies on the NSG not only to
  search for neighbors of an agent, but also to accurately determine the agents
  in a specific subvolume for agent migrations, aura updates, and load
  balancing.
To avoid the high overheads of rebuilding the complete NSG after each of these steps,
we adapted the implementation to allow for incremental changes, i.e., the addition, removal, and position update of single agents.

\head{Modularity Improvements}
Alongside high performance, modularity is another key design aspect of \ta{}.
We have therefore made additional efforts in this direction during the
  development of \ta{} and made three main changes.
First, we introduce the \texttt{VisualizationProvider} interface to facilitate
  rendering of additional information besides agents and scalar/vector fields.
We use an implementation of this interface to render the partitioning grid as
  can be seen in Figure~\ref{fig:eval:validation}.
Second, we add the \texttt{SimulationSpace} interface to explicitly gather
  information about whole and local simulation space in one place.
Third, we introduce the
  \texttt{SpaceBoundaryCondition} interface and refactor the implementations
  for ``open'', ``closed'', and ``toroidal'' boundaries.

\section{Evaluation Methodology}
\label{sec:evaluation_methodology}

\subsection{Experimental Setup and Reproducibility}
All tests were executed in a Singularity container with an Ubuntu 22.04-based
  image.
We use two systems to evaluate the distributed simulation engine, demonstrating \ta{} on both a high-end cluster and a commodity setup.
First, we use the Dutch National Supercomputer Snellius Genoa partition, also referred as the System A,
  where each node has two AMD Genoa 9654 processors with 96 physical CPU cores
  each, 384 GB total memory per node, and an InfiniBand interconnect (200 Gbps
within a rack, 100 Gbps outside the rack) \cite{snellius_hardware}.
Second, we use a two-node system (System~B) where each node has four Intel Xeon E7-8890~v3 processors with 72 physical CPU cores per node in total, 504\,GB of main memory on the first node (used for the single-node studies in \secref{sec:de:eval:bdm} and \secref{sec:de:eval:interoperability}) and 1024\,GB on the second, and a Gigabit Ethernet interconnect.
We measure the ``simulation runtime'' in terms of the wall-clock time needed to simulate a number of iterations, which excludes the initialization step and teardown.
Each experiment is repeated three times; we report the mean, with error bars showing the standard deviation, unless stated otherwise.
We will provide all code and instructions to execute the benchmarks on Zenodo upon publication.

\subsection{Benchmark Simulations}
\label{sec:benchmarks}
We test \ta{} with four of the five simulations that the state-of-the-art work~\cite{breitwieser_biodynamo_2023, breitwieser-bdm} uses for performance evaluation: cell clustering (Biocellion variant), cell proliferation, and use cases in the domains of epidemiology and oncology. Table~\ref{tab:sim-characteristics} shows that these simulations cover a broad spectrum of performance-relevant characteristics.
We exclude the neuroscience use case because its agents modify their neighbors, which \ta{} does not yet support: it would require agents to merge partial results and an additional distributed communication step.
\begin{table}
  \centering
  \footnotesize
  \caption{Performance-relevant characteristics of the benchmark simulations (\secref{sec:benchmarks}).}
\begin{tabular}{@{}lrrrr@{}}
    \toprule
    \textbf{Characteristic} & 
    \rotatebox{90}{\textbf{Cell proliferation}} & 
    \rotatebox{90}{\textbf{Cell clustering}} & 
    \rotatebox{90}{\textbf{Epidemiology}} & 
\rotatebox{90}{\textbf{Oncology}} \\
    \midrule
    {Create new agents} &
    \cmark  &
      &
      &
\cmark  \\

    {Delete agents} &
      &
      &
      &
\cmark  \\

    {Load imbalance} &
      &
      &
\cmark &
      \\

    {Agents move randomly} &
      &
      &
    \cmark  &
\cmark  \\

    \bottomrule
  \end{tabular}
\label{tab:sim-characteristics}
\end{table}

\section{TeraAgent Functional Evaluation}
\label{sec:functional_evaluation}

\subsection{Correctness}

\begin{figure}[bth]
  \centering
  \begin{subfigure}[b]{.48\linewidth}
    \centering
    \includegraphics[width=\linewidth]{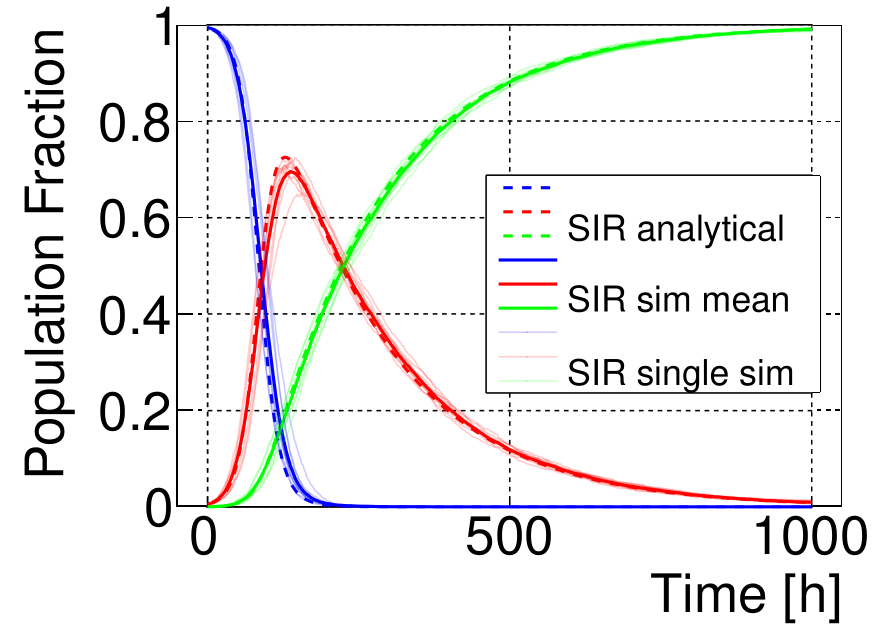}
    \caption{}
    \label{fig:eval:validation:epidemiology}
  \end{subfigure}
  \hfill
  \begin{subfigure}[b]{.48\linewidth}
    \centering
    \includegraphics[width=\linewidth]{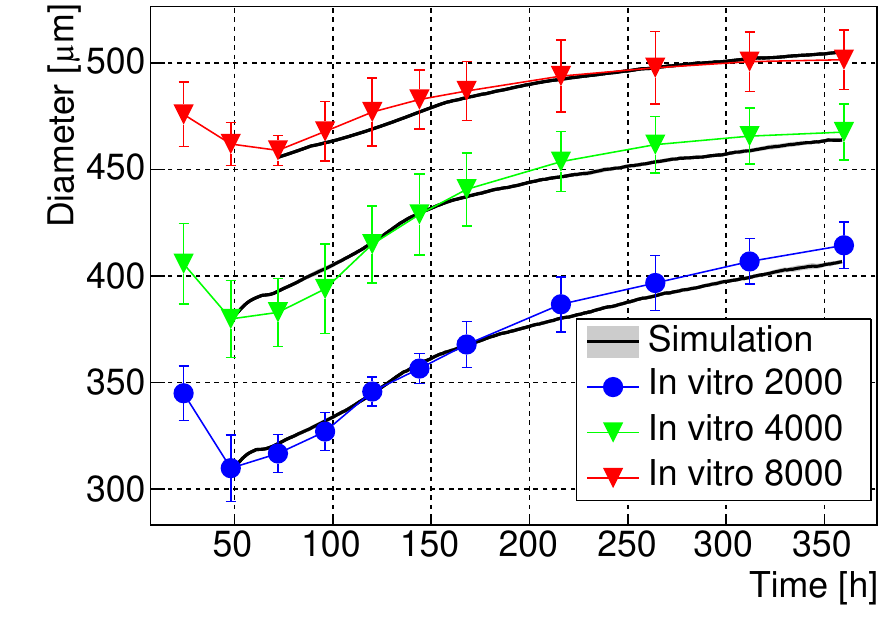}
    \caption{}
    \label{fig:eval:validation:tumor}
  \end{subfigure}

  \vspace{1em}

  \begin{subfigure}[b]{.4\linewidth}
    \centering
    \includegraphics[width=\linewidth]{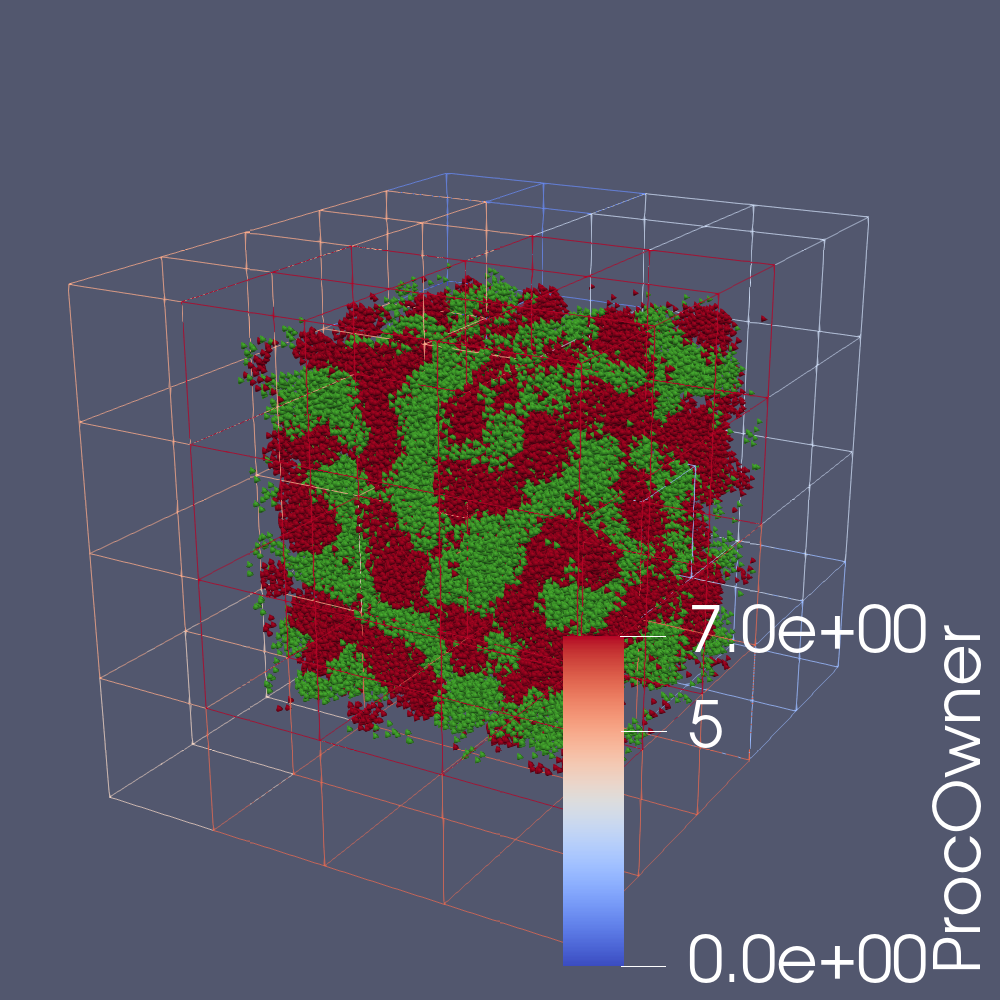}
    \caption{}
    \label{fig:eval:validation:clustering}
  \end{subfigure}

  \caption{Result verification of \ta{}. (a)~Epidemiology and (b)~oncology use cases. (c)~Cell clustering, final state on eight MPI ranks illustrating the space partitioning grid colored by the rank that owns the volume (\texttt{ProcOwner})}
\label{fig:eval:validation}
\end{figure}

To ensure the correctness of the distributed simulation engine, we replicate the results of prior work~\cite{breitwieser-bdm, breitwieser_biodynamo_2023} and add 180 automated unit and integration tests.
\figref{fig:eval:validation} compares \ta{}'s results against the analytical SIR solution (epidemiology use case, \figref{fig:eval:validation:epidemiology}), against \emph{in vitro} measurements of tumor spheroid growth (oncology use case, \figref{fig:eval:validation:tumor}), and shows the final outcome of the cell clustering simulation (\figref{fig:eval:validation:clustering}).
We observe that \ta{}'s curves match both references across the whole simulated time span, reproducing the results that \bdm{} obtains for the same models~\cite{breitwieser-bdm, breitwieser_biodynamo_2023}, and qualitatively replicating the cell clustering results of \cite{breitwieser_biodynamo_2023, kang2014biocellion}.

\subsection{Seamless Transition From a Laptop to a Supercomputer}
\label{sec:eval:laptop-to-sc}

User-friendliness is a key design aspect, alongside our focus on performance. With regard to distributed computing, the model definitions for the four benchmark simulations are completely transparent to the user. Only the evaluations for the epidemiology and oncology use cases require additional code for distributed execution.

The changes in the epidemiology simulation are limited to \emph{two} lines of
  code.
To create the plot shown in \figref{fig:eval:validation:epidemiology}, the engine has
  to count the number of agents for the three groups (susceptible, infected, and
  recovered).
For distributed executions, the rank-local results have to be summed up to
  obtain the correct result.
To do that, we provide the function \texttt{SumOverAllRanks}, which hides the
  MPI-related function call.
Furthermore, to save the generated plot to disk, we have to make sure that only
  one rank writes to the same file location.
This goal can be achieved by telling the other ranks to exit the function
  before the save function is executed.
We can achieve this with the preprocessor macro
  \texttt{IF\_NOT\_RANK0\_RETURN}.

The tumor spheroid simulation also requires extra code to generate the result
  plot (\figref{fig:eval:validation:tumor}).
To accurately measure the tumor diameter in the simulation, we determine the
  volume of the convex hull, from which we can calculate the diameter by assuming
  a spherical shape.
\ifthesis
Chapter~\ref{ch:engine} uses libqhull \cite{libqhull}, a
  library which is not distributed.
\else
Breitwieser \etal{} \cite{breitwieser-bdm} used libqhull \cite{libqhull}, a
  library which is not distributed.
\fi
Consequently, the simulation contains a couple of lines of extra code to transmit
  agent positions to the root rank to perform the diameter calculation.
For simulations with a larger number of agents, we use a more approximate
  method by determining the enclosing bounding box.
The approximate method is provided by \ta{} and is the same
  whether or not the simulation is distributed.
We conclude that \ta{} users do \emph{not} need to know about
  distributed computing to scale out a simulation to tens of thousands of CPU
  cores. 
Researchers can start the development on a laptop and seamlessly transition to
  more powerful hardware as the model size and complexity grow.

\section{TeraAgent's Scalability}
\label{sec:scalability}
\label{sec:de:eval:scalability}

We evaluate \ta{}'s scalability in two different case studies.

\head{Case Study: Strong and Weak Scaling on Snellius}
We test two complementary scaling strategies: strong scaling (reducing time-to-result for fixed problem size) and weak scaling (maintaining efficiency as problem size grows proportionally with resources). For strong scaling, we fix the problem size per simulation at the largest model that saturates a single Snellius node (0.5--1.2 billion agents, depending on the simulation) and execute 10 iterations on 1 to 16 nodes (3,072 CPU cores), increasing by powers of two.

For weak scaling\footnotemark, we fix the problem size at 100 million agents per node and scale from 1 to 128 nodes (24,576 CPU cores), scaling up the number of agents to 12.8 billion. 
\footnotetext{Results for the oncology use case are omitted from the weak scaling analysis due to  an unexpected runtime error encountered, which could not be resolved within the allocated computational budget.}
Both scaling studies use global load balancing with Zoltan2's recursive coordinate bisection (RCB).

\begin{figure}[h!]
    \centering
    \includegraphics[width=.65\linewidth]{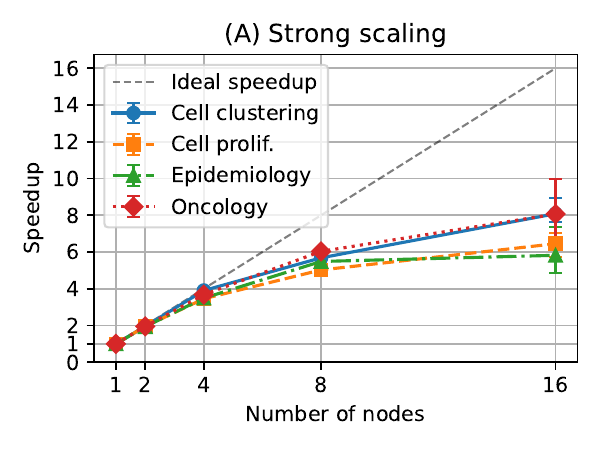}
    \includegraphics[width=.65\linewidth]{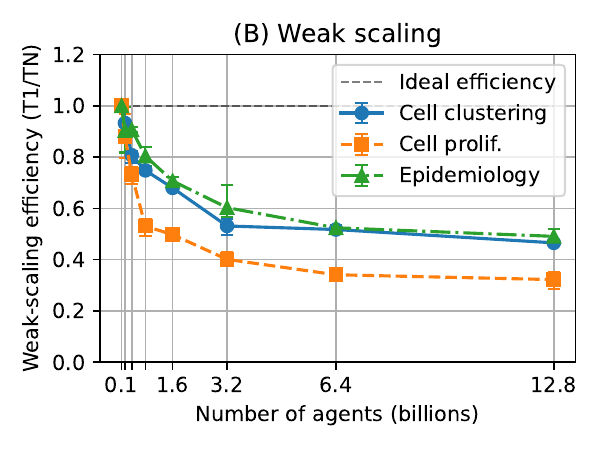}
	\caption{Scalability of \ta{} on Snellius: (A)~strong scaling and (B)~weak scaling. The gray dashed line marks ideal behavior in both panels. The error bars represent the minimum and maximum observed values.}
	\label{fig:eval:strong-scaling}
\end{figure}

\figref{fig:eval:strong-scaling}A and~\ref{fig:eval:strong-scaling}B show the strong- and weak-scaling analyses, respectively.
\figref{fig:eval:strong-scaling}A shows the effect of scaling the simulation across more nodes (x-axis) on the performance speedup (y-axis) for a constant problem size. Therefore, as the number of nodes increases on the x-axis, each node simulates a smaller number of agents. The ideal speedup is proportional to the number of nodes, as annotated with a gray dashed line.
\figref{fig:eval:strong-scaling}B shows the effect of increasing both the problem size (i.e., number of agents) and the number of nodes on the weak-scaling efficiency (y-axis), i.e., the single-node runtime divided by the runtime at the given node count. Regardless of the number of nodes, each node simulates 100 million agents in this scalability study. We annotate the ideal scenario as a gray dashed line. The ideal trend is at $y=1.0$ because efficiency should ideally remain at 100\% when the number of nodes and the number of agents increase at the same rate.

We make two observations from \figref{fig:eval:strong-scaling}. 
First, \ta{}'s strong scaling up to 4 nodes (768 CPU cores) follows an almost linear scalability curve, retaining 86--98\% parallel efficiency and thus demonstrating an effective work distribution. Efficiency drops to 36--50\% at 16 nodes (3,072 CPU cores), which we attribute to load imbalances and increased synchronization overhead.
Second, weak-scaling efficiency declines gradually with scale, reaching 32--49\% at 128 nodes: simulating a $128\times$ larger model on $128\times$ more nodes increases the runtime per iteration by only $2.0$--$3.1\times$, confirming \ta{}'s ability to handle proportionally larger problems with added resources.
Based on these observations, we conclude that \ta{} practically scales to both reduce time-to-result (strong scaling) and enable larger simulations (weak scaling).

\head{Case Study: Extreme Scaling on Snellius}
To observe \ta{}'s scalability at an unprecedentedly large number of agents, we conduct two additional tests. 
First, we execute a simulation with \ESMidAgents{} agents for 10 iterations on \ESMidNumNodes{} nodes (\ESMidNumCPUs{} CPU cores).

Second, we increase the number of agents even further to \emph{\ESMaxAgentsExact{}} and use \ESNumNodes{} nodes with \ESNumCPUs{} CPU cores, 
the maximum number of cores that we were permitted to use on the Dutch National Supercomputer. To fit this many agents into the available main memory, we reduce the engine's memory consumption by disabling all optimizations that require extra memory: using single-precision floating-point numbers, reducing the agent's size by changing the base class, and reducing the memory consumption of the neighbor search grid.
We observe that both simulations completed 10 iterations successfully. The simulation with \ESMidAgents{} agents consumed \ESMidMemoryConsumptionInTB{}~TB of memory and took \ESMidRuntimePerSecond{} seconds per iteration, while the simulation with \ESMaxAgentsExact{} agents consumed \ESMemoryConsumptionInTB{}~TB of memory and took \ESRuntimePerSecond{} seconds per iteration.
These test results show that \ta{} can successfully simulate \ESMaxAgentsInTrillion{} agents: a $\ESFactorImprovementSOTA\times$ improvement over the largest simulation demonstrated with \bdm{}~\cite{breitwieser_biodynamo_2023} and Biocellion~\cite{kang2014biocellion} (\ESAgentsBDM{} agents), and an $\ESFactorImprovementLiterature\times$ improvement over the largest agent count reported in the literature (\ESAgentsLiterature{} agents~\cite{parker_2007}).
These two runs are capability demonstrations rather than scaling measurements: they establish that \ta{} can hold and advance a model of this size, whereas \figref{fig:eval:strong-scaling} characterizes how efficiently it scales.
\ta{} can in principle scale even more. Unfortunately, we could \emph{not} test larger ABMs due to a lack of hardware resources. We hope and expect that our insights and our implementation, which we will open-source, will enable both future research on improving simulation efficiency and further investment in computational resources for extreme-scale agent-based simulations.

\section{Comparison with BioDynaMo}
\label{sec:de:eval:bdm}
\begin{figure}[h]
  \centering
	\includegraphics[width=0.8\linewidth]{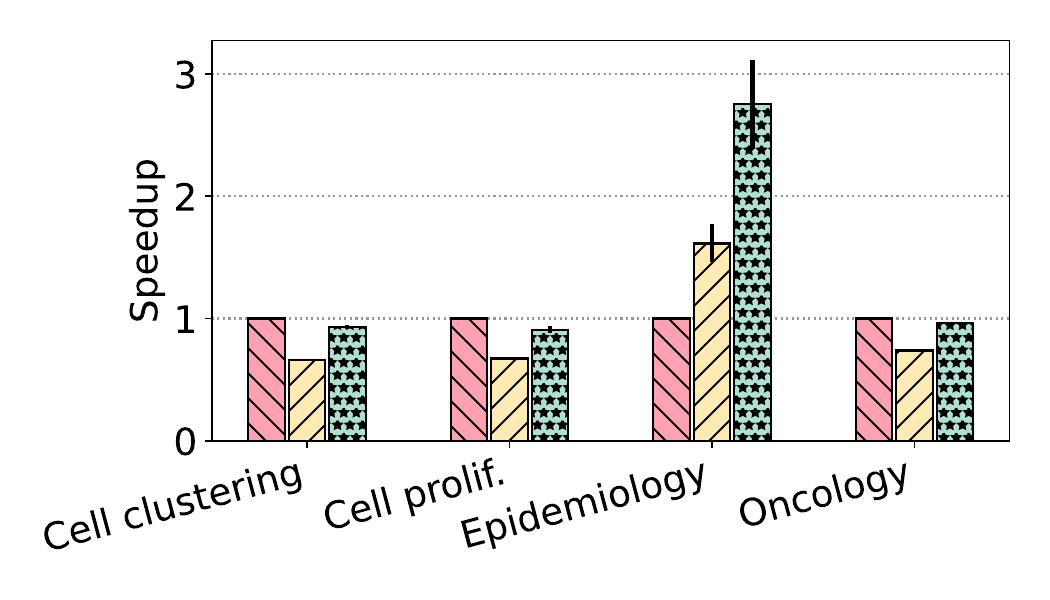}\\
	\includegraphics[width=0.8\linewidth]{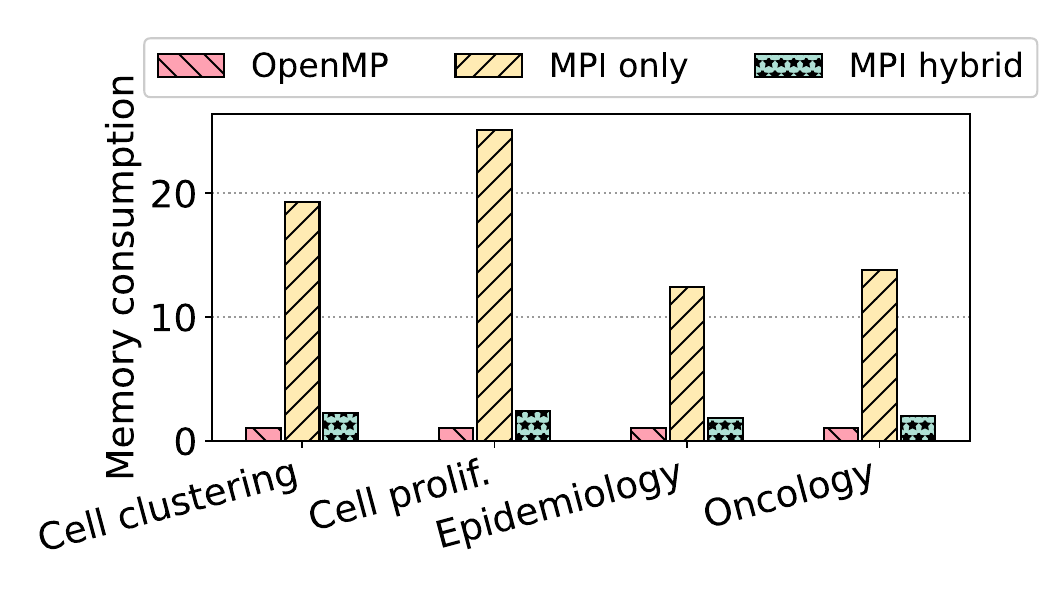}
  \caption{Speedup (top) and normalized memory consumption (bottom) with respect to \bdm{}. Each group shows three bars: OpenMP (the \bdm{} baseline, normalized to 1), MPI-only, and MPI-hybrid. The legend is shared between the plots.}
	\label{fig:eval:mpi-vs-openmp}
\end{figure}

We evaluate \ta{}'s performance for two configurations, MPI-only and MPI-hybrid, and compare it against the performance of \bdm{}, which is OpenMP-only. 
The benchmarks were executed on one node of System~B with $10^7$ agents and for
  all iterations of the simulations.
\figref{fig:eval:mpi-vs-openmp} shows our performance and memory consumption results on the top and bottom subplots, respectively. The x-axes on both subplots show the simulation and the y-axes are normalized to \bdm{}'s results.

We make two major observations from \figref{fig:eval:mpi-vs-openmp}. 
First, MPI-hybrid mode comes within 4--9\% of OpenMP (\bdm{}) on three of the four simulations while adding distributed execution capability, and it is always better than MPI-only mode. On the epidemiology use case, MPI-hybrid instead \emph{outperforms} OpenMP by $2.8\times$.
We attribute the epidemiology speedup to reduced cross-CPU traffic, {inline with a prior work's~\cite{breitwieser-bdm} observation } 
that this simulation's performance is particularly sensitive to the memory layout, and although \bdm{} is NUMA-aware, not all of its data structures (e.g., the neighbor search grid) are replicated per NUMA domain. \ta{}'s MPI-hybrid mode launches one rank per NUMA domain, which partitions these structures and outweighs the overheads of aura updates and agent migrations. More generally, MPI-hybrid mode combines 1)~high-performance parallel execution using OpenMP within a rank and 2)~scalable parallelism using MPI across ranks.

Second, across all tested simulations, OpenMP achieves the lowest memory consumption by leveraging its shared-memory-based design, whereas MPI-only has the highest, averaging $17.6\times$ that of OpenMP. While MPI increases memory consumption by more than an order of magnitude, \ta{}'s MPI-hybrid approach has a memory consumption of $2.1\times$ that of OpenMP across tested simulations. 
Therefore, MPI-hybrid's memory overhead above the OpenMP baseline is \emph{only} 6.6\% of MPI-only's.
We attribute MPI-only's memory consumption overhead to the use of the C++ interpreter cling~\cite{vasilev_cling_2012}, because each rank instantiates its own cling interpreter, which occupies several hundred megabytes. These results show that \ta{}'s MPI-hybrid design significantly alleviates MPI's high memory consumption. 

We conclude that \ta{}'s MPI-hybrid approach achieves scalable high performance at a low memory overhead.

\subsection{Improved Interoperability}
\label{sec:de:eval:interoperability}

\head{Motivation}
Relevant third-party software achieves scalability primarily through MPI rather than shared-memory parallelism.
\bdm{}'s OpenMP-only architecture creates a fundamental interoperability barrier: these tools cannot effectively distribute work when communicating with a single-rank application. \ta{}'s MPI support removes this barrier, enabling practical integration with the broader scientific computing ecosystem.

\begin{figure}[h!]
  \centering
	\includegraphics[width=.65\linewidth]{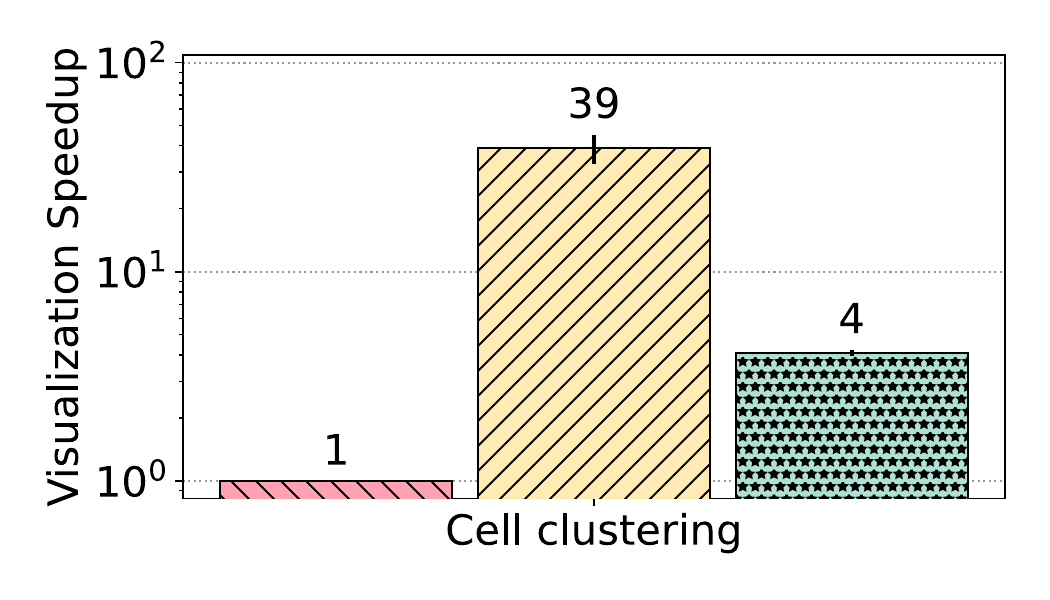}
	\includegraphics[width=.65\linewidth]{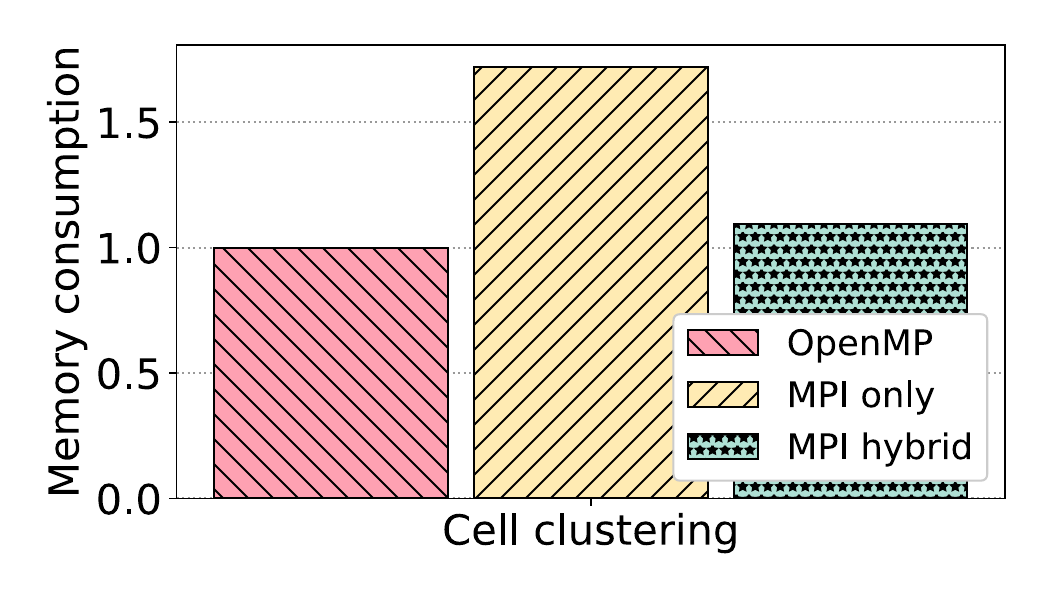}
	\caption{In-situ visualization with ParaView on System~B: visualization speedup (top) and normalized memory consumption (bottom) for \ta{} in MPI-only and MPI-hybrid configurations, both with respect to \bdm{} (OpenMP). The legend is shared between the plots.}
	\label{fig:eval:insitu-performance}
\end{figure}

\head{Case Study: ParaView In-Situ Visualization}
We demonstrate this improvement using ParaView, \bdm{}'s primary visualization tool~\cite{breitwieser_biodynamo_2023}. ParaView offers two modes: export mode, which writes simulation state to disk for later visualization, and in-situ mode, which visualizes data directly from memory during simulation.  
For this demonstration, we execute the cell clustering simulation on System~B with 10 million agents for 10 iterations, each of which renders a separate image. We compare the visualization performance and the memory consumption of three different configurations: OpenMP (\bdm{} baseline) with 144 threads on one rank, MPI-only with a thread in each of 72 ranks, and MPI-hybrid with 36 threads in each of 4 ranks.
MPI-only uses 72 threads rather than 144 because it launches one rank per \emph{physical} CPU core and therefore does \emph{not} use hyperthreading, \emph{unlike} the other two configurations.
\figref{fig:eval:insitu-performance} shows visualization performance (top) and memory consumption (bottom). In both subplots, the x-axis specifies the simulation, the three bars in each group correspond to the three configurations, and the y-axes show the respective results normalized to the \bdm{} baseline.

We make three observations from \figref{fig:eval:insitu-performance}.
First, the MPI-only configuration achieves $39\times$ higher visualization performance than OpenMP despite using half the computational resources (72 vs.\ 144 threads). This dramatic improvement occurs because ParaView's visualization workload scales with the number of MPI ranks available for parallel execution: \bdm{}'s single-rank architecture severely limits ParaView's parallelism, while \ta{}'s 72 ranks enable effective work distribution.
Second, MPI-hybrid with 4 ranks achieves $4\times$ the visualization performance of OpenMP on a single rank, despite these two configurations having the same number of threads (i.e., 144 threads).
This result shows that visualization performance can significantly benefit from increasing the number of MPI ranks rather than the number of threads in a rank. 
Third, {the memory consumption values of MPI-only and MPI-hybrid are 72.1\% and 9.5\% larger than OpenMP, respectively. We attribute this increase to: 1) each rank} maintaining its own cling instance (several hundred MB per rank), and 2) replication of distributed simulation data structures (partitioning grid metadata, communication buffers). However, because ParaView's visualization data structures dominate total memory consumption, these per-rank overheads result in relatively modest increases. 
Based on these observations, we conclude that \ta{}'s MPI architecture unlocks practical in-situ visualization by enabling ParaView to distribute work across multiple ranks.
We expect other MPI-parallelized tools such as the lattice Boltzmann solver OpenLB~\cite{krauseOpenLBOpenSource2021} to benefit similarly, although we have not evaluated this. \ta{} therefore expands both the achievable simulation scale and the range of third-party tools that can be integrated into agent-based simulation workflows.

\section{Comparison with BioCellion}
\label{sec:de:eval:biocellion}

\head{Motivation} 
Biocellion is a distributed agent-based simulation platform specialized for tissue modeling that reported simulating 1.72 billion agents (i.e., the largest agent count in prior work besides epidemiological models).

\head{Evaluation} We execute the cell clustering simulation with 1.72 billion cells on System~B using two nodes (144 physical CPU cores total). 
We measure runtime averaged over all iterations and compute the agent update rate per CPU core as the performance metric. For Biocellion, we use the results published by Kang \etal{}~\cite{kang2014biocellion}, as the software is \emph{not} available under an open-source license. Biocellion executed the same simulation on 4,096 CPU cores (AMD Opteron 6271 Interlagos) with 4.46s per iteration.

Table~\ref{tab:biocellion_comparison} shows the comparison between \ta{} and Biocellion. 

\begin{table}[h]
\centering
\caption{Comparison of \ta{} and Biocellion on the cell clustering simulation with 1.72 billion agents.}
\label{tab:biocellion_comparison}
\small
\begin{tabular}{lrr}
\toprule
Metric & \ta{} & Biocellion \\
\midrule
Runtime per iteration & 15.8\,s & 4.46\,s \\
CPU cores & 144 & 4,096 \\
Number of agents & 1.72B & 1.72B \\
Agent updates per sec. per core & $7.56 \times 10^5$ & $9.42 \times 10^4$ \\
Open source & Planned & No\\
\bottomrule
\end{tabular}
\end{table}

We make two observations. First, despite \ta{}'s longer absolute runtime per iteration (15.8\,s vs.\ 4.46\,s), \ta{} achieves 8$\times$ higher computational efficiency per CPU core.
Second, Biocellion is \emph{not} available under an open-source license, whereas we will release \ta{} as open source.
Based on these observations, we conclude that \ta{} significantly advances the state-of-the-art in distributed agent-based simulations, with an $8\times$ higher agent update rate per CPU core.

\section{Improvement Breakdown}
\subsection{Serialization}
\label{sec:de:eval:serialization}

\begin{figure}
	\centering
\begin{subfigure}{\linewidth}
    \centering
    \includegraphics[width=0.75\textwidth]{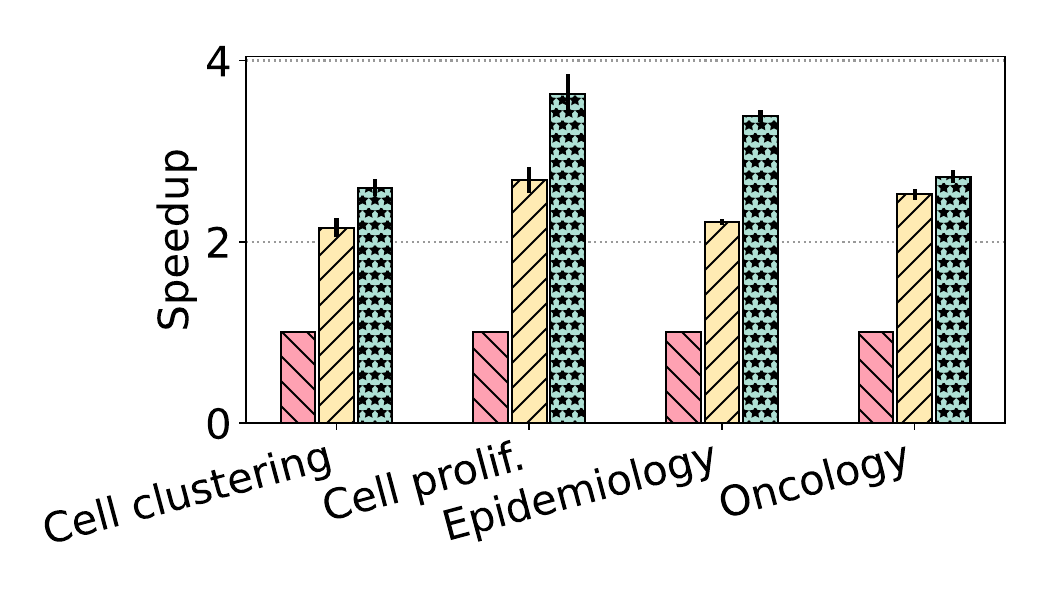}
    \includegraphics[width=0.75\textwidth]{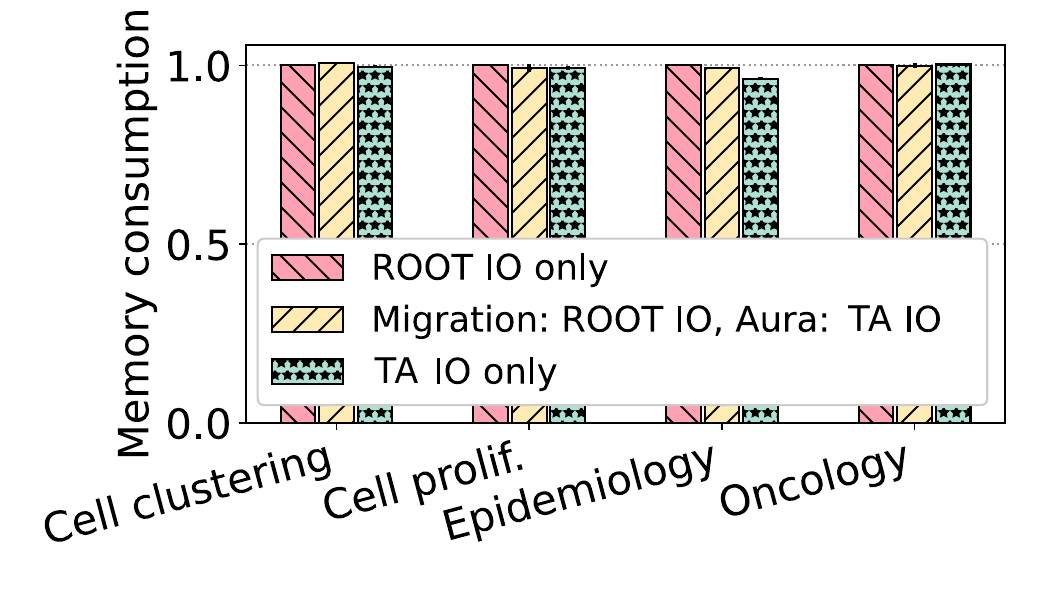}
\caption{Overall simulation speedup (top) and memory consumption (bottom) compared to ROOT IO, including both exclusive \ta{}-IO and a hybrid configuration (ROOT IO for migrations and \ta{}-IO for aura updates). The legend is shared between the two plots.}
    \label{fig:eval:serialization:overall}
   \end{subfigure}
  \begin{subfigure}{\linewidth}
    \centering
    \vspace{5mm}
    \includegraphics[width=0.49\textwidth]{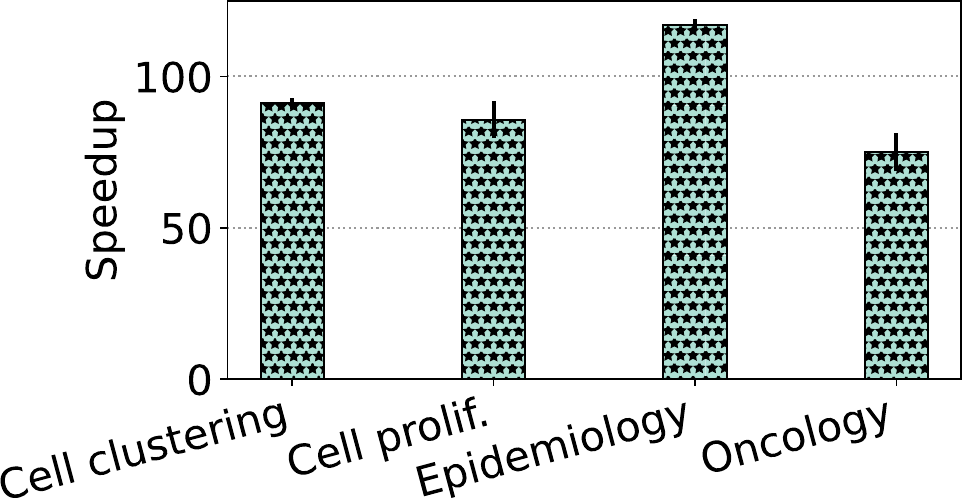}
    \includegraphics[width=0.49\textwidth]{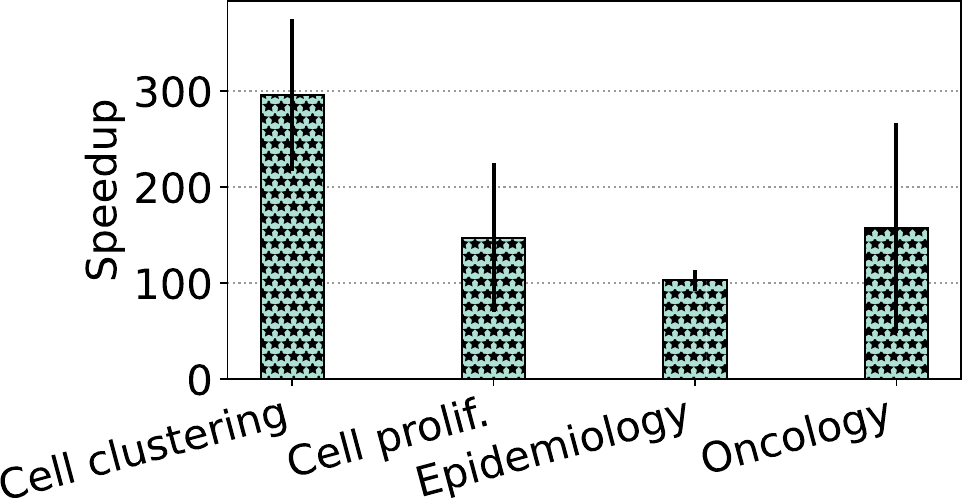}
\caption{\ta{}-IO serialization speedup relative to ROOT IO for aura updates (left) and agent migrations (right).}
    \label{fig:eval:serialization:serialization}
   \end{subfigure}
  \begin{subfigure}{\linewidth}
    \centering
    \vspace{5mm}
    \includegraphics[width=0.49\textwidth]{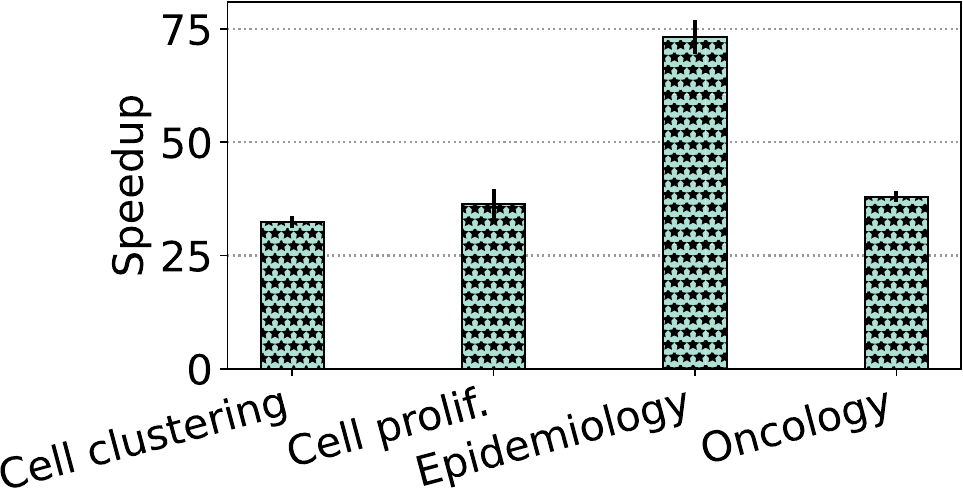}
    \includegraphics[width=0.49\textwidth]{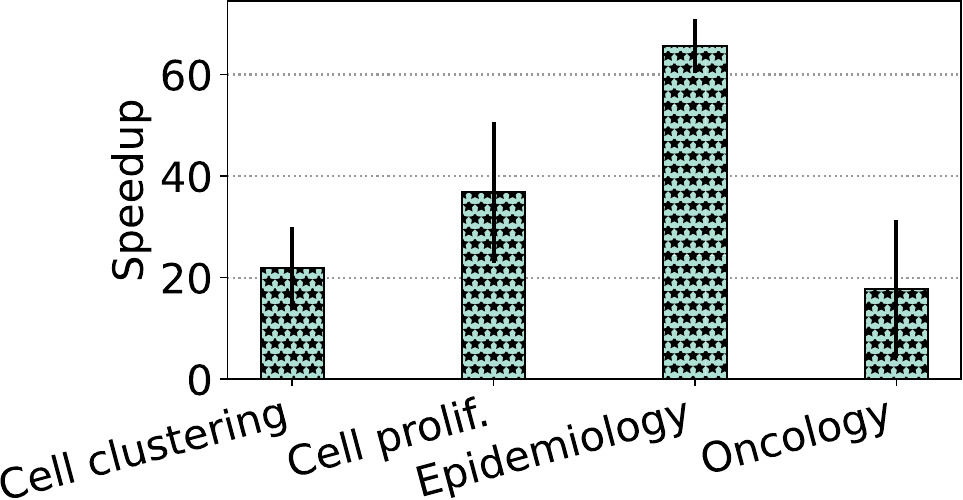}
    \caption{\ta{}-IO deserialization speedup relative to ROOT IO for aura updates (left) and agent migrations (right).}
    \label{fig:eval:serialization:deserialization}
   \end{subfigure}
  \begin{subfigure}{\linewidth}
    \centering
    \includegraphics[width=\textwidth]{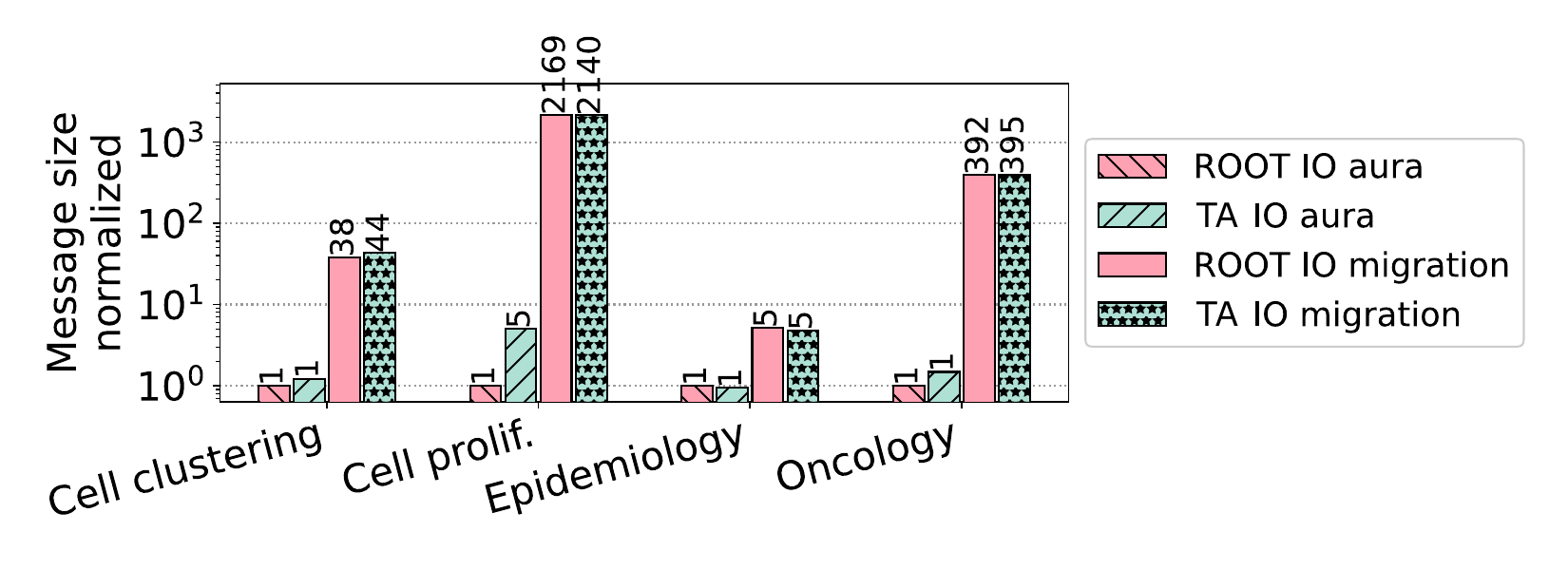}
    \caption{Comparison of \ta{} and ROOT IO message sizes, normalized to the size of ROOT IO aura messages.}
    \label{fig:eval:serialization:message-size}
   \end{subfigure}
\caption{Comparison of the \ta{}-IO serialization mechanism (labeled \texttt{TA IO} in the plots) and ROOT IO on Snellius.}

	\label{fig:eval:serialization}
\end{figure}

To evaluate the performance of \ta{}'s serialization mechanism compared to ROOT IO, we perform benchmark simulations on four Snellius nodes in MPI-hybrid mode using $10^8$ agents over 10 iterations. We measure simulation runtime and memory consumption, serialization and deserialization speedups, and message sizes (\figref{fig:eval:serialization}).
Our results show that simulation runtime is reduced
by up to 3.6$\times$ while memory consumption remains constant (\figref{fig:eval:serialization:overall}).
\ta{}-IO serializes the agents up to \result{296$\times$} faster (median \result{110$\times$}, \figref{fig:eval:serialization:serialization}) and also significantly improves the deserialization performance, with a maximum observed speedup of \result{73$\times$} (median \result{37$\times$}, \figref{fig:eval:serialization:deserialization}).
We verify that the resulting message sizes are equivalent across \ta{}-IO and ROOT IO (\figref{fig:eval:serialization:message-size}). 
The only outlier, in cell proliferation, is due to its small message size and
  does not negatively impact performance.

\subsection{Data Transfer Minimization}
\label{sec:de:eval:delta-encoding}

\begin{figure}[tbh!]
	\centering
\begin{subfigure}{\linewidth}
    \centering
    \includegraphics[width=0.65\textwidth]{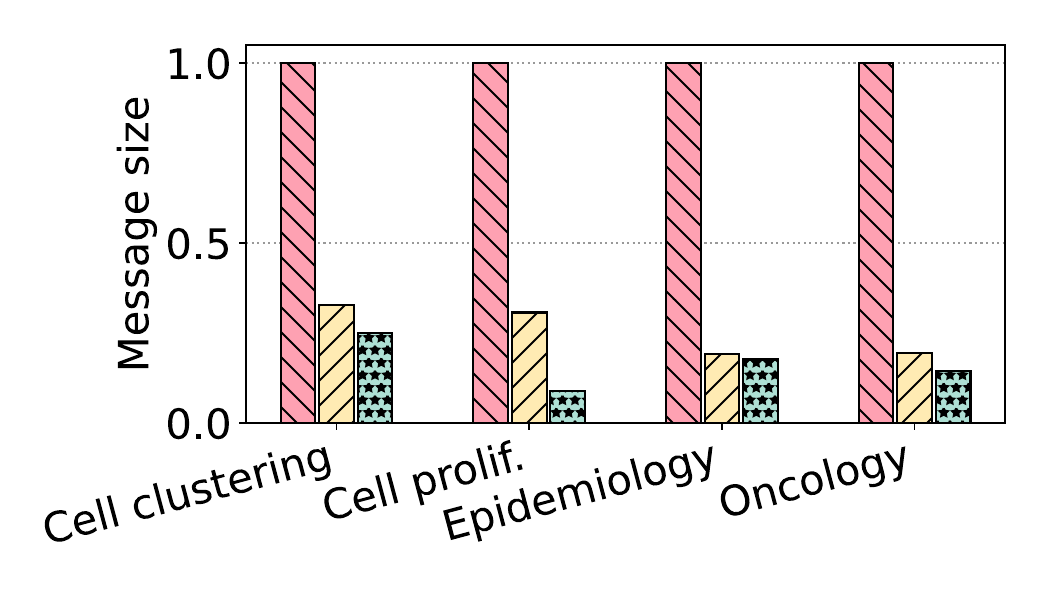}
    \caption{Normalized message size per configuration.}
    \label{fig:eval:serialization-compression:message-size}
   \end{subfigure}
  \begin{subfigure}{\linewidth}
    \centering
    \includegraphics[width=0.65\textwidth]{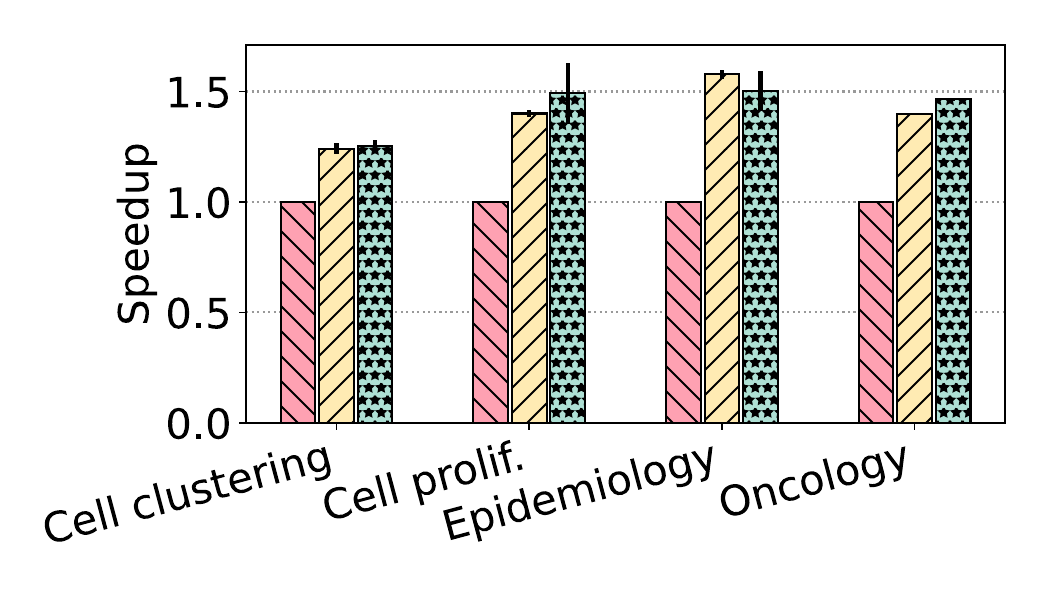}
    \caption{Overall simulation speedup.}
    \label{fig:eval:serialization-compression:speedup}
   \end{subfigure}
  \begin{subfigure}{\linewidth}
    \centering
    \includegraphics[width=0.65\textwidth]{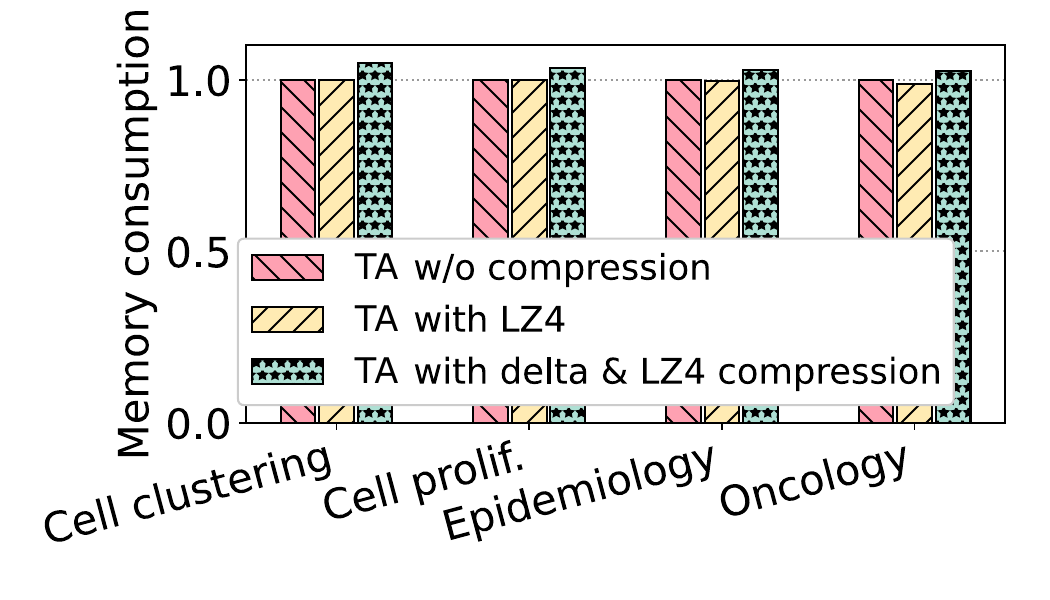}
    \caption{Memory consumption.}
    \label{fig:eval:serialization-compression:memory}
   \end{subfigure}
  \begin{subfigure}{\linewidth}
    \centering
    \includegraphics[width=0.65\textwidth]{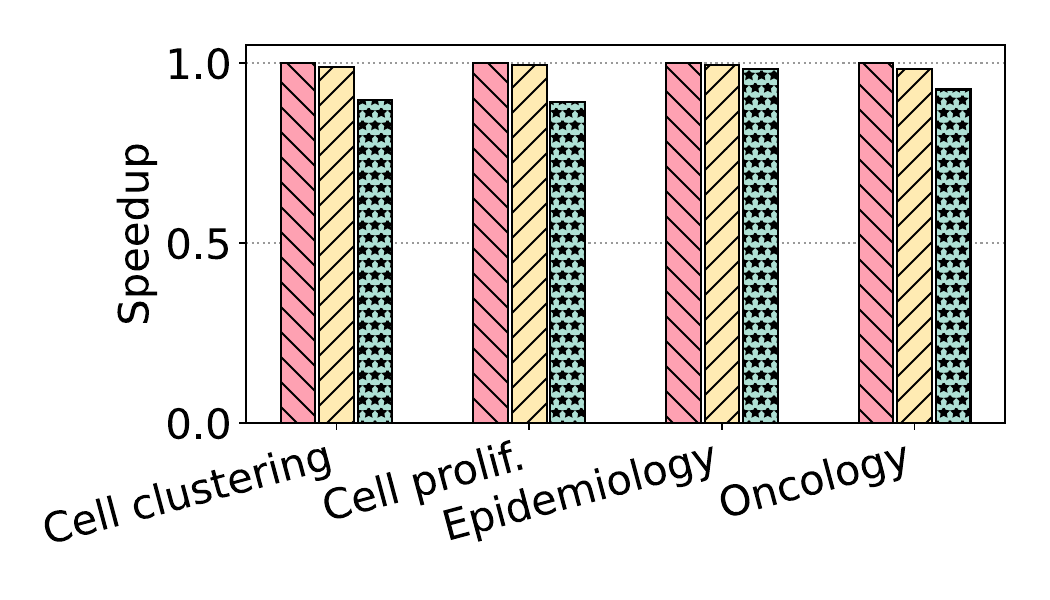}
    \caption{Speedup of distribution operation during a single iteration.}
    \label{fig:eval:serialization-compression:distribution-op}
   \end{subfigure}
  \begin{subfigure}{\linewidth}
    \centering
    \includegraphics[width=0.65\textwidth]{./fig-24}
    \caption{Speedup of agent update operation during a single iteration.}
    \label{fig:eval:serialization-compression:agent-op}
   \end{subfigure}
\caption{Effect of compression on System~B. Each group shows three bars: no compression (the \ta{}-IO baseline, normalized to 1), LZ4 only, and LZ4 with delta encoding. The legend is shared among all plots.}
	\label{fig:eval:serialization-compression}
\end{figure}

To evaluate the performance improvements of LZ4 compression \cite{lz4} and delta encoding, we execute all benchmark simulations with $10^8$ agents on two System~B nodes and four Snellius nodes for 10 iterations.
We observe that the message size is reduced by \result{3.0--5.2}$\times$ by LZ4 compression and by a further \result{1.1--3.5}$\times$ when adding the delta encoding scheme described in \secref{sec:design:delta-encoding}.
This improvement speeds up the distribution operation, which subsumes aura updates and agent migrations, by up to \result{11}$\times$.
However, the significant speedups of delta encoding do \emph{not} directly translate to the entire execution of the simulation.
\figref{fig:eval:serialization-compression} shows the performance of two compression configurations (LZ4 only, and LZ4 with delta encoding) on System~B, both normalized to the \ta{}-IO baseline with \emph{no} compression.
Specifically, we show the reduction in message size (\figref{fig:eval:serialization-compression:message-size}), the impact on overall runtime (\figref{fig:eval:serialization-compression:speedup}) and memory usage (\figref{fig:eval:serialization-compression:memory}), and the speedup of distribution and agent update operations (\figref{fig:eval:serialization-compression:distribution-op}~and~\ref{fig:eval:serialization-compression:agent-op}).

We make three observations from \figref{fig:eval:serialization-compression}.
First, adding delta encoding on top of LZ4 compression yields a small additional end-to-end gain on System~B: $+7\%$ for cell proliferation and $+5\%$ for oncology, $+1\%$ for cell clustering (within the error bars), and $-4\%$ for epidemiology (\figref{fig:eval:serialization-compression:speedup}). 
We attribute the gains to the reduced message size, and thus faster communication across ranks.
Second, delta encoding reduces the performance of agent operations by 2--12\% (\figref{fig:eval:serialization-compression:agent-op}), which we attribute to the agent reordering step.
Third, we observe that the memory consumption increases slightly (by 3\% on average across simulations) due to the reference copies kept on each rank (\figref{fig:eval:serialization-compression:memory}).

In general, the benefit of delta encoding depends strongly on the interconnect. On System~B (Gigabit Ethernet), LZ4 compression alone reduces simulation runtime for all simulations and delta encoding adds up to a further $7\%$. On Snellius (200~Gbps InfiniBand), however, the reduced message size does \emph{not} translate into runtime savings, because the encoding overhead outweighs the transfer savings.

Based on these results, we conclude that data movement can be a major contributor to the overall simulation time, and that our compression scheme is most valuable on commodity and cloud interconnects (i.e., precisely the deployments that motivate \ta{}'s hardware-flexibility goal).
 
\section{Related Work}
\label{sec:related-work}
\ta{} significantly improves the performance and the scalability of agent-based simulations by leveraging insights into the application domain.

\head{Agent-Based Simulation Tools}
To our knowledge, \ta{} is the only simulation platform capable of simulating \ESMaxAgentsExact{} agents.
The largest reported agent populations in the literature are from Parker~\cite{parker_2007}, who simulated a specialized epidemiological model with up to \ESAgentsLiterature{} agents, and Biocellion~\cite{kang2014biocellion}, a distributed tool in the tissue modeling domain, with \ESAgentsBDM{} agents.
We discuss the remaining related (distributed) platforms below, each
characterized by how it parallelizes and the largest scale it reports.

Repast HPC~\cite{repasthpc} distributes agents as C++ objects across MPI
processes and provides projections (grid, network, and continuous space) that 
 define the agents' environment. 
Repast HPC's benchmark was run on an
IBM Blue Gene/P system using up to 32,768 of its 40,960 processors, reaching
a maximum of $10^9$ agents.
The paper attributes the resulting bottleneck to
a root rank that serially coordinates agent-sharing requests.

Cytowski and Szymańska~\cite{cytowski_large-scale_2014,cytowski_large-scale_2015}
use a hybrid MPI+OpenMP parallelization strategy for cell-based agent
models.
Their evaluation on an IBM Blue Gene/Q system used 512 nodes
(32,768 hardware threads), also reaching but not exceeding $10^9$ agents.
This agent count is notably of the same order of
magnitude as Repast HPC's, despite the later hardware generation. Unlike the
Repast HPC paper, their work~\cite{cytowski_large-scale_2014,cytowski_large-scale_2015} does
not describe a specific bottleneck limiting further scale.

D-MASON~\cite{dmason,cordascoDistributedMASONScalable2018} is written in Java and
partitions the simulation into cells, each owned by a logical process.
Exchange with neighboring processes is either performed centrally over JMS/ActiveMQ
or decentralized with MPI.
Its largest
reported on-premises run reaches $\sim$20 million agents on a 15-node cluster with 180 cores.
A separate cloud deployment on AWS, which could in principle scale elastically, was tested
only up to 144 cores.
The paper's own stated contributions concern correctness and usability rather
than raw scale.

Chaste~\cite{chaste} parallelizes its continuum cardiac-tissue PDE solver via
PETSc and (par)METIS, with independent benchmarks~\cite{strazdinsPerformanceAnalysisCardiac2011}
reporting scaling to 2,048 cores.
However, the paper does not report a comparable scalability evaluation for its
  cell-based, agent-style simulation component.

FLAME GPU~\cite{richmond_high_2010} and its successor FLAME GPU 2~\cite{richmondFLAMEGPU22023} both employ a single-simulation-per-GPU execution model, which limits the maximum number of agents to the capacity of the GPU's memory. 
Richmond \etal{} have demonstrated a Sugarscape model~\cite{epsteinGrowingArtificialSocieties1996} with up to 16 million agents. 
FLAME GPU 2's distributed capabilities are limited to executing ensemble simulations, i.e., running multiple instances of the same model to improve statistical confidence in stochastic results or to perform parameter exploration. 
This method is orthogonal to the single simulation distributed across multiple compute nodes presented in this paper.

To summarize, the platforms above had access to substantial computing
resources ranging from an 80\%-utilized 40,960-processor Blue Gene/P to an
elastically scalable cloud deployment, yet no platform has demonstrated
simulations at the scale reported by \ta{}.

\head{Serialization}
There is a wide range of serialization libraries.
\ifthesis
Chapter~\ref{ch:dse}
\else
This paper
\fi
compares the performance against ROOT IO \cite{brun_root_1997},
  which according to Blomer \cite{blomer_quantitative_2018} outperforms
  Protobuf \cite{protobuf}, HDF5 \cite{hdf5}, Parquet \cite{parquet}, and Avro
  \cite{avro}.
MPI \cite{mpi} also provides functionality to define derived data types, but targets use cases with regular patterns, for example, the last element of each row in a matrix. \ta{}'s agents are allocated on the heap with irregular offsets between them and, therefore, cannot use MPI's solution.

\head{Delta Encoding}
Delta encoding \cite{hunt_algorithm_1976} is a widely used concept to minimize the amount of data that is
  stored or transferred, which we apply to aura updates of the agent-based
  workload (\secref{sec:design:delta-encoding}).
Other applications include backups \cite{burns1997efficient}, file revision
  systems such as git \cite{git}, network protocols \cite{rfc1144,
	  mogul2002delta}, cache and memory compression \cite{pekhimenko_base-delta-immediate_2012, pekhimenko_linearly_2013, lee_adaptive-latency_2015, vijaykumar_case_2015}, and more \cite{macdonald2000file, delta_encoding_vms}.
We did not find any explicit mention in the literature of using this concept to accelerate the distributed
  execution of agent-based simulations.

\section{Discussion and Future Work}
\label{sec:discussion_and_future_work}

Although \ta{} enables simulating half-trillion-scale agent-based models on distributed systems, we acknowledge several limitations.
First, our experimental evaluation is constrained by the computational resources we were granted, and a more comprehensive scaling study at intermediate node counts would further characterize performance bottlenecks. 
Second, accelerator-based execution (e.g., GPUs and processing-in-memory) is a promising direction given the inherently parallel nature of ABMs, but the pointer-rich object hierarchies, load imbalance, control-flow divergence, and data structure impedance 

  pose significant challenges for such architectures; we leave this to future work.
Third, we focus our evaluation on the distributed communication layer (i.e., the primary bottleneck when scaling out) rather than single-node characterization, which has been studied in prior work~\cite{breitwieser-bdm, breitwieser_biodynamo_2023} and remains applicable as \ta{} preserves \bdm{}'s intra-node execution strategy.

\section{Conclusion}

We present \ta{}, the most scalable distributed simulation engine for agent-based models (ABMs), which we will release as open source.

\ta{} implements a new data serialization mechanism and a new data compression scheme that collectively reduce the data movement across distributed partitions of an agent-based model and provide the partitions with a native interface to communicate with third-party software.   
By doing so, \ta{} 1)~enables extreme-scale simulations with \ESMaxAgentsInTrillion{} agents, 2)~reduces time-to-result by adding compute nodes, 3)~improves interoperability with third-party tools, and 4)~provides users with greater hardware flexibility.

\ta{} allows researchers to scale out their execution environment seamlessly from laptops and workstations to clouds and supercomputers.
We demonstrate 
that in many use cases, such scale-out does \emph{not} require any model code changes.
We hope and expect that \ta{} will inspire research in various domains that can benefit from high-performance and scalable agent-based simulations by enabling researchers to generate new insights into complex systems. To foster future research, we plan to open-source \ta{} with the publication of this paper.

\section*{Acknowledgment}
This work used the Dutch National Supercomputer Snellius, with the support of SURF Cooperative under grant no. EINF-5667.
We also want to thank CERN openlab for providing access to two servers.
We acknowledge the support provided to the SAFARI Research Group by our
  industrial partners, including Huawei, Intel, Microsoft, and VMware.

\bibliographystyle{IEEEtran}

\begin{IEEEbiography}[{\includegraphics[width=1in,height=1.25in,clip,keepaspectratio]{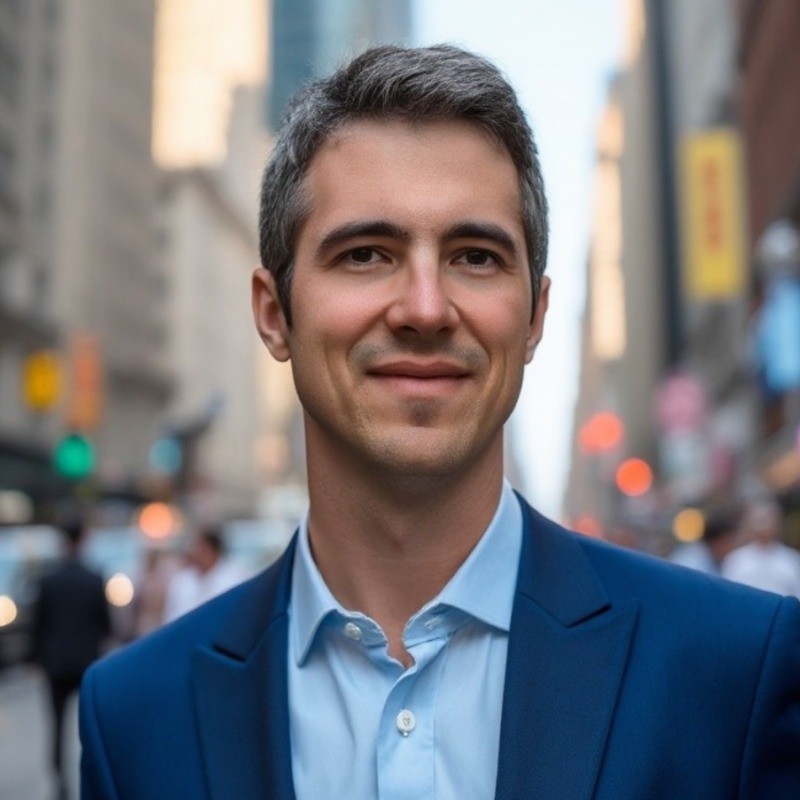}}]{Lukas Breitwieser}
received his doctoral degree in computer science from ETH Zurich, Switzerland, in 2024, under the supervision of Prof. Onur Mutlu (SAFARI Research Group) and Dr. Fons Rademakers (CERN openlab), and holds an MSc in software development and business management from Graz University of Technology. 
His research interests encompass high-performance, parallel, distributed, and heterogeneous computing, with applications in large-scale simulation, data analysis, and AI.
Through his work at the European Organization for Nuclear Research (CERN) and ETH Zurich, he has addressed fundamental limitations in agent-based modeling regarding performance, efficiency, scale, and modularity. 
Through the development of the \bdm{} and \ta{} platforms, he has enabled researchers to study complex systems at unprecedented scales across diverse scientific domains; notably, a radiotherapy model based on \bdm{} was recognized as one of the top 10 breakthroughs in physics for 2024.
In recognition of these contributions, Dr. Breitwieser received the Impact Award from the Society of Research Software Engineering in the UK in 2025 and the Best Artifact Award from the Symposium on Principles and Practice of Parallel Programming (PPoPP) in 2023. 
At CERN, he currently focuses on accelerating high-energy physics data analysis through GPU-based computing.
\end{IEEEbiography}

\begin{IEEEbiography}
[{\includegraphics[width=1in,height=1.25in,clip,keepaspectratio]{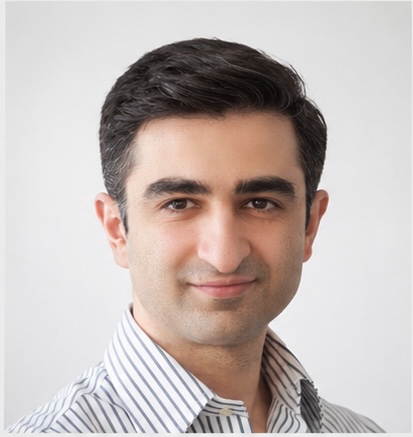}}]{Ahmad Hesam}
received his MSc in computer engineering from the Delft University of Technology (the Netherlands) in 2018.
From 2019 to 2023, he was a software engineer at the European
Organization for Nuclear Research (CERN), with a specialized focus on GPU acceleration for scientific applications.
He is currently a Solutions Architect at the Dutch national research and education organisation (SURF), where he designs system-level architectures for research infrastructure and scientific workflows.
\end{IEEEbiography}

\begin{IEEEbiography}[{\includegraphics[width=1in,height=1.25in,clip,keepaspectratio]{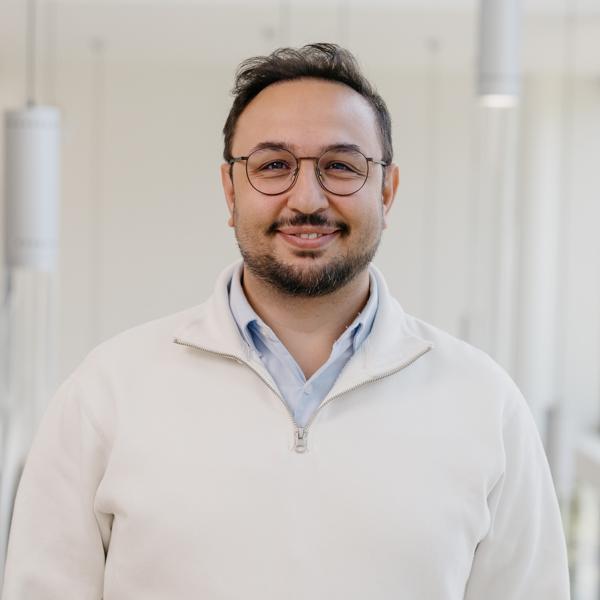}}]{ABDULLAH GIRAY YA\u{G}L{I}K{\c{C}}{I}} is a tenure-track faculty at CISPA. His broader research interests span high-performance, energy-efficient, and secure computer architectures aiming robustly and sustainably scalable systems. His research is published in major venues, including HPCA, MICRO, DSN, ISCA, and USENIX Security. He received his Dr. sc. degree from ETH Zurich, SAFARI Research Group, in 2024, advised by Prof. Onur Mutlu. His dissertation builds 1) a detailed understanding of DRAM read disturbance, a major limitation of main memory density scaling, and 2) mechanisms that efficiently and scalably mitigate DRAM read disturbance. His research was awarded with ACM SIGMICRO Dissertation Award and William C. Carter PhD Dissertation Award in Dependability in 2025 and received an honorable mention by ACM SIGARCH/IEEE CS TCCA Outstanding Dissertation Award. His research is in part supported by Google Security and Privacy Research Award and Microsoft Swiss Joint Research Center. Contact him at giray@cispa.de.
\end{IEEEbiography}

\begin{IEEEbiography}[{\includegraphics[width=1in,height=1.25in,clip,keepaspectratio]{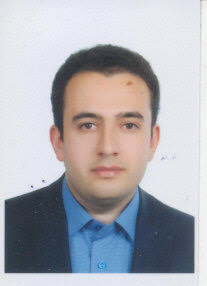}}]{Mohammad Sadrosadati}
received his Ph.D. degree in Computer Engineering from Sharif University of Technology, Tehran, Iran, in 2019 under the supervision of Prof. H. Sarbazi-Azad. He spent one year from April 2017 to April 2018 as an academic guest at ETH Zurich hosted by Prof. O. Mutlu during his Ph.D. program. He is currently a senior researcher in the SAFARI research group at ETH Zurich. His research interests are in the areas of heterogeneous computing, processing-in-memory, memory systems, and interconnection networks. Due to his achievements and impact on improving the energy efficiency of GPUs, he won Khwarizmi Youth Award, one of the most prestigious awards, as the first laureate in 2020, to honor and embolden him to keep taking even bigger steps in his research career.
\end{IEEEbiography}

\begin{IEEEbiography}[{\includegraphics[width=1in,height=1.25in,clip,keepaspectratio]{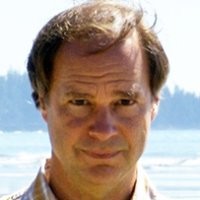}}]{Fons Rademakers}
received his Ph.D. degree in particle physics from the University of Amsterdam, the Netherlands, in 1991, for his work on event displays and data analysis for the DELPHI experiment at CERN’s LEP collider.
He has been with the European Organization for Nuclear Research (CERN) since 1991, where he first worked on the PAW project and developed column-wise ntuples and PIAF, an early parallel and distributed data analysis system. 
In 1995, he started the ROOT (root.cern) project together with René Brun, and he served as its project leader from 2011 to 2014, through the release of ROOT 6.
ROOT is used to store, process, and analyze the data of particle physics experiments worldwide. In 2014 he moved on to CERN openlab, first as Chief Technology Officer and subsequently as Chief Research Officer.
In 2015 he co-founded the \bdm{} (biodynamo.org) project, an open-source agent-based simulator for complex biological systems, medicine development, epidemiology, finance, animal behavior, and socio-economic issues.
He served as Technical Coordinator for \bdm{} until the release of v1.04 in late 2022.
In 2025 he reached the CERN mandatory retirement age, but stays involved as a CERN Honorary Associate with the ROOT project.
\end{IEEEbiography}

\begin{IEEEbiography}[{\includegraphics[width=1in,height=1.25in,clip,keepaspectratio]{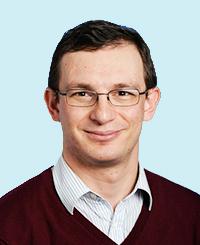}}]{Onur Mutlu} {\,}  (Fellow, IEEE and ACM) is currently a Professor of Computer Science at ETH Zurich. He is also a Faculty Member at Carnegie Mellon University, where he previously held the Strecker Early Career Professorship. His research interests include computer architecture, computing systems design, hardware security, and bioinformatics. Various techniques he, along with his group and collaborators, has invented have influenced industry and are employed in commercial microprocessors, memory systems, and storage systems. He started the Computer Architecture Group, Microsoft Research, from 2006 to 2009, and held various product, research and visiting positions at Intel Corporation, AMD, VMware, Google, and Stanford University. He received various honors for his research, including the 2025 IEEE Computer Society Harry H. Goode Memorial Award ``for seminal contributions to computer architecture research and practice, especially in memory systems'' as well as five Test of Time Awards from major computer architecture and systems conferences. He is an AAAS Fellow, ACM Fellow and an elected member of the Academy of Europe. His computer architecture and digital design course lectures are freely available (https://www.youtube.com/OnurMutluLectures). His research group makes a wide variety of software and hardware artifacts freely available online (https://safari.ethz.ch/). Onur Mutlu received his Ph.D. degree in computer engineering from the University of Texas at Austin and Bachelor's degrees in computer engineering and psychology from the University of Michigan, Ann Arbor. He is a Fellow of IEEE, Fellow of the Association for Computing Machinery, and a Fellow of the American Association for the Advancement of Science. Contact him at omutlu@gmail.com.
\end{IEEEbiography}
 
\EOD
\end{document}